\documentclass[twoside,onecolumn]{article}
\usepackage{blindtext} 

\usepackage[sc]{mathpazo} 
\usepackage[T1]{fontenc} 
\usepackage{microtype} 
\usepackage[english]{babel} 
\usepackage[hmarginratio=1:1,top=32mm,columnsep=20pt]{geometry} 
\usepackage[hang, small,labelfont=bf,up,textfont=it,up]{caption} 
\usepackage{booktabs} 
\usepackage{lettrine} 
\usepackage{enumitem} 
\setlist[itemize]{noitemsep} 
\usepackage{abstract} 
\usepackage{titlesec} 
\renewcommand\thesection{\Roman{section}} 
\renewcommand\thesubsection{\roman{subsection}} 
\titleformat{\section}[block]{\large\scshape\centering}{\thesection.}{1em}{} 
\titleformat{\subsection}[block]{\large}{\thesubsection.}{1em}{} 
\usepackage{fancyhdr} 
\usepackage{graphics}
\usepackage{graphicx}
\usepackage{epsfig}
\usepackage{amssymb}
\usepackage{cite}
\usepackage{titling} 
\usepackage{hyperref} 
\usepackage{ragged2e}
\usepackage{amsmath}
\usepackage{mathptmx}
\usepackage{graphicx}  
\usepackage{dcolumn}   
\usepackage{bm}        
\usepackage{amssymb}   
\usepackage{subfigure}  
\usepackage{hyperref}
\usepackage{ragged2e}
\usepackage{mathtools}
\usepackage{xcolor}
\usepackage{nccmath}
\usepackage{ragged2e}
\usepackage{graphics}
\usepackage{graphicx}
\usepackage{epsfig}
\usepackage{amssymb}
\usepackage{lineno}
\usepackage{blindtext}
\usepackage{lineno}
\usepackage{lineno, blindtext}
\usepackage{mathtools}
\usepackage{xcolor}
\usepackage{hyperref}
\usepackage{microtype}
\usepackage{breakcites}
\usepackage{footnote}
\usepackage{appendix}
\usepackage{ragged2e}

\pretitle{\begin{center}\LARGE\bfseries} 
	\posttitle{\end{center}} 
\title{Experimental and numerical evidence of intensified non-linearity at the micro and nano scale: The lipid coated acoustic bubble} 
\author{%
	\textsc{AJ. Sojahrood $^{a,b}$} \thanks{Corresponding author email: amin.jafarisojahrood@ryerson.ca},\textsc{H. Haghi$^{a,b}$}, \textsc{T.M. Porter$^c$}  \\[1ex] %
	\textsc{ R. Karshafian$^{a,b}$}\textsc{Michael C. Kolios $^{a,b}$}\\
	\normalsize $^a$ Department of Physics, Ryerson University, Toronto, Ontario, Canada. \\
	\normalsize $^b$ Institute for Biomedical Engineering and Science Technology, \\
	\normalsize A Partnership Between Ryerson University and St. Michael’s Hospital, Toronto, Canada. \\
	\normalsize $^c$ Department of Biomedical Engineering, The University of Texas at Austin, Texas, USA\\
 }
\date{\today} 
\renewcommand{\maketitlehookd}{
	\begin{abstract}
			A lipid coated bubble (LCB) oscillator is a  very interesting non-smooth oscillator with many important applications ranging from industry and chemistry to medicine. However, due to the complex behavior of the coating intermixed with the nonlinear behavior of the bubble itself, the dynamics of the LCB are not well understood. In this work, lipid coated Definity$^{\textregistered}$ microbubbles (MBs)  were sonicated with 25 MHz 30 cycle pulses with pressure amplitudes between 70kPa-300kPa. Here, we report  higher order subharmonics in the scattered signals of single MBs at low amplitude high frequency ultrasound excitations. Experimental observations reveal the generation of period 2(P2), P3, and two different P4 oscillations at low excitation amplitude. Despite the reduced damping of the uncoated bubble system, such enhanced nonlinear oscillations has not been observed and can not be theoretically explained for the uncoated bubble. To investigate the mechanism of the enhanced nonlinearity, the bifurcation structure of the lipid coated  MBs is studied for a wide range of MBs sizes and shell parameters. Consistent with the experimental results, we show that this unique oscillator can exhibit chaotic oscillations and higher order subharmonics at excitation amplitudes considerably below those predicted by the uncoated  oscillator. Buckling or rupture of the shell and the dynamic variation of the shell elasticity causes the intensified non-linearity at low excitations. The simulated scattered pressure by single MBs are in good agreement with the experimental signals.  
	\end{abstract}
}

\begin{document}

		\maketitle

		\section{Introduction}
	\justify
Acoustically excited bubbles are known to be nonlinear oscillators \cite{A,B} with "infinite complexity" \cite{A}. The addition of a lipid stabilizing shell increases this complexity, thus, even after over a decade of study, the dynamics of ultrasonically excited lipid coated microbubbles (MBs)  are not fully understood. Interestingly, lipid coated MBs have been shown to exhibit 1/2 order subharmonic (SH) oscillations even when the excitation amplitude is low ($<$30 kPa \cite{1,2,3}). Despite the increased inherent damping due to the coating, such low threshold values contradict the predictions of the theoretical models as these values are even below the thresholds expected for uncoated free MBs \cite{4,5}.\\ 
The lipid coating may cause compression dominated oscillations \cite{6} or limit the MB oscillations to only above a certain pressure threshold \cite{7}. It has been shown in \cite{1} that the low pressure threshold for 1/2 order SH emissions is due to the compression only behavior of the MBs due to the buckling of the shell. Overvelde et al.\cite{8} showed that the lipid coating may enhance the nonlinear MB response at acoustic pressures \textit{as low as 10 kPa}. In addition, even a small ($\approx10 kPa$) increase in the acoustic pressure amplitude leads to a significant decrease in the main resonance frequency \cite{8} resulting to a pronounced skewness of the resonance curve.  The origin of the “thresholding” \cite{7} behavior has been linked to the shift in resonance \cite{8}. Nonlinear resonance behavior of the lipid coating has also been observed at higher frequencies (5-15 MHz\cite{9}), (8-12 MHz \cite{10}) and (11-25 MHz \cite{11}). Through theoretical analysis of the Marmottant model for lipid coated MBs \cite{12}, Prosperetti \cite{4} attributed the lower SH threshold of the lipid MBs to the variation in the mechanical properties of the coating in the neighborhood of a certain MB radius (e.g. the occurrence of buckling). The coating may also result in expansion dominated behavior in liposome-loaded MBs \cite{13}.  Expansion dominated oscillations occur when the initial surface tension of the lipid coated MB is close to that of the water \cite{11,13}. In this regime, and in contrast to compression dominated behavior, the MB expands more than it compresses. Expansion-dominated behavior was used to explain the enhanced non-linearity at higher frequencies (25 MHz) \cite{11}. The Marmottant model effectively captures the behavior of the MB including expansion-dominated behavior \cite{11,12,14}, compression only behavior \cite{6}, thresholding \cite{7}  and enhanced non-linear oscillations at low excitation pressures \cite{1,2,8,14,15,16}.  Despite the numerous studies which employed the methods of nonlinear dynamics and chaos to investigate the dynamics of acoustically excited MBs \cite{A,B,20,21,22,23,24,25,26,27,28,29,30,31,32,33,34,35,36,37,38,39,40,41,42,43,44}, the detailed bifurcation structure and nonlinear dynamics of the lipid coated MBs have not been studied. \\ Lipid coated MBs are being routinely used in diagnostic ultrasound \cite{2,45,46,47,48}. Moreover, they have been used in pioneering non-invasive treatments of brain disorders and tumors in humans \cite{49}. Currently there are several investigations on the potential use of lipid coated micro-bubbles (MBs) in high resolution and high contrast imaging procedures \cite{50} as well as non-invasive ultrasonic treatments and localized drug/gene delivery \cite{51}. Despite the promising results of these investigations, the complex dynamics of the system makes it difficult to optimize the behavior of lipid coated MBs. Moreover, from the nonlinear dynamics point of view, the lipid coated MB oscillator is an interesting topic of investigation due to the highly nonlinear nature of the system. The complex behavior of the uncoated MB is interwoven with the nonlinear behavior of the lipid coating which enables unique dynamical properties for this oscillator. \\ 
In this work we study the bifurcation structure of the lipid coated MBs as a function of size and frequency at different pressure values. Numerical results are accompanied by experimental observations of single MB scattering events at low pressure excitation. We show for the first time, both experimentally and numerically that in addition to 1/2 order SHs, higher order SHs (e.g. 1/3, 1/4) can be generated at low excitation amplitudes. Analysis of the bifurcation structure of the system reveals a unique property of the lipid coated MB which is the generation of chaos at excitation pressures as low as 5kPa.
\section{Methods}
\subsection{Experimental method}
Very dilute solutions of Definity$^{\textregistered}$ MBs were sonicated with 25 MHz continuous pulse trains using the Vevo 770 ultrasound imaging machine (VisualSonics Inc. Toronto, Ontario). The pulse length was held constant at 30 cycles while the applied acoustic pressure was varied over the range of $\approx 70-300 kPa$. The backscattered signals from individual MBs were extracted and different nonlinear modes of oscillations were identified. Acquisition of signals from single MB were similar to the approach in \cite{52}. Fig. 1 shows a schematic of an acquired signal from a single MB event ($\approx250 kPa$ and $f=25 MHz$).
\begin{figure*}
	\begin{center}
		\includegraphics[scale=3]{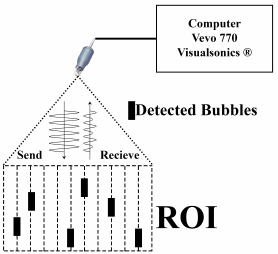}\\
		(a)
		\includegraphics[scale=0.5]{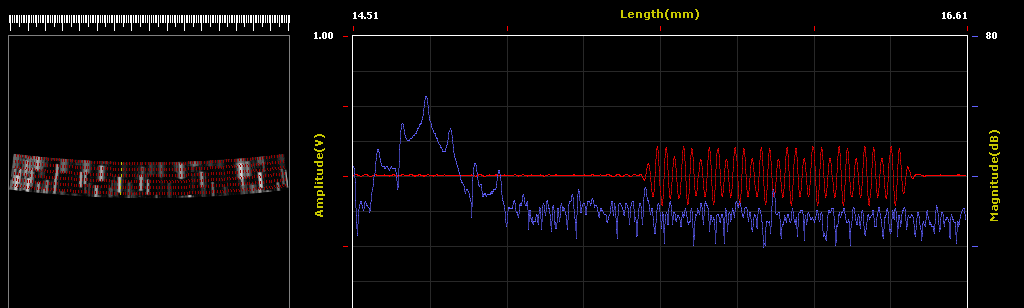}\\
		(b)
		\caption{a) Schematic of the Vevo 770 (Visualsonics$^{\textregistered}$) machine which was used in the experiments to detect the signals from single MB events in the region of interest (ROI). b) Left: ROI selected for an ultrasound pulse train at 25 MHz and 250 kPa of pressure (each large subdivision is $\approx$ 130 $\mu m$), and Right: Signal (red) from a single period-3 MB event. The frequency spectrum of the received signal is shown in blue exhibiting 1/3 order SHs at 8.33 and	 2/3 order SHs 16.66 MHz.}
	\end{center}
\end{figure*}
\subsection{Numerical procedure}
\subsubsection{Marmottant Model}
Dynamics of coated MBs undergoing buckling and rupture can be effectively modeled using the Marmottont equation \cite{12}:
\begin{linenomath*}
	\begin{equation}
	\begin{gathered}
	\rho \left(R\ddot{R}+\frac{3}{2}\dot{R}^2\right)=\\\left[P_0+\frac{2\sigma(R_0)}{R_0}\right](\frac{R}{R_0})^{-3k}\left(1-\frac{3k}{c}\dot{R}\right)- P_0-\frac{2\sigma(R)}{R}-\frac{4\mu\dot{R}}{R}-\frac{4k_s\dot{R}}{R^2}-P_a(t)
	\end{gathered}
	\end{equation}
\end{linenomath*}
In this equation, R is radius at time t, $R_0$ is the initial MB radius, $\dot{R}$ is the wall velocity of the bubble, $\ddot{R}$ is the wall acceleration,	$\rho{}$ is the liquid density (998 $\frac{kg}{m^3}$), c is the sound speed (1481 m/s), $P_0$ is the atmospheric pressure, $\sigma(R)$ is the surface tension at radius R, $\mu{}$ is the liquid viscosity (0.001 Pa.s), $k_s$ is the coating viscosity and  $P_a(t)$ is the amplitude of the acoustic excitation ($P_a(t)=P_a sin(2\pi f t)$) where $P_a$ and \textit{f} are the amplitude and frequency of the applied acoustic pressure. The numerical values in the parentheses are for pure water at 293 K. The gas inside the MB is C3F8 and water is the host medium.\\ 
The surface tension $\sigma(R)$ is a function of radius and is given by:
\begin{linenomath*}
	\begin{equation}
	\sigma(R)=
	\begin{dcases}
	0  \hspace{3cm}   if \hspace{1cm}  R \leq R_b\\
	\chi(\frac{R^2}{R_b^2}-1) \hspace{1.2cm} if\hspace{1cm}  R_b \leq R\leq R_{r}\\
	\sigma_{water} \hspace{2cm}  if\hspace{.5cm} \hspace{.5cm}  R > R_r
	\end{dcases}
	\end{equation}
\end{linenomath*}
$\sigma_{water}$ is the water surface tension (0.072 N/m), $R_b=\frac{R_0}{\sqrt{1+\frac{\sigma_0}{\chi}}}$ is the buckling radius, $R_r=R_b\sqrt{1+\frac{\sigma_r}{\chi}}$ is the rupture radius,  $\chi$ is the shell elasticity, $\sigma_0$ is the initial surface tension at $R=R_0$, and $\sigma_r$ is the rupture surface tension. Similar to \cite{11} in this paper $\sigma_r=\sigma_{water}$. Shear thinning of the coating is included in the Marmottant model using \cite{53}:
\begin{linenomath*}
	\begin{equation}
	k_s=\frac{4k_0}{1+\alpha \frac{|\dot{R}|}{R}};
	\end{equation}
\end{linenomath*}
where $k_0$ is the shell viscous parameter and $\alpha$  is the characteristic time constant associated with the shear rate. In this work shell parameters of $\chi=0.975 N/m$, $k_0=2.98\times 10^{-10} kg s^{-1}$ and $\alpha=1\times 10^{-6} s$  are used for the $Definity^{\textregistered}$ MBs. These values are adopted from the estimated parameters in \cite{53,54,Carly, Qian}.\\ Due to the negligible thermal damping of C3F8 even at high amplitude oscillations \cite{US2,US3} thermal damping is neglected in this paper.
\subsubsection{Keller-Miksis model}
The dynamics of the uncoated MBs were also visualized alongside the lipid coated MBs to highlight the contributions of the coating to the nonlinear bheavior of the bubble. To model the uncoated MB dynamics the Keller-Miksis model \cite{55} is used:
\begin{linenomath*}
	\begin{equation}
	\rho[(1-\frac{\dot{R}}{c})R\ddot{R}+\frac{3}{2}\dot{R}(1-\frac{R}{3c})]=(1+\frac{\dot{R}}{c})(G)+\frac{R}{c}\frac{d}{dt}(G)
	\label{eq:5}
	\end{equation}
\end{linenomath*}
where $G=\left( P_0+\frac{2\sigma_{water}}{R_0} \right)(\frac{R}{R_0})^{-3k}-\frac{4\mu_L\dot{R}}{R}-\frac{2\sigma}{R}-P_0-P_A sin(2 \pi f t)$.\\
\subsubsection{Scattered pressure by MBs}
The pressure scattered (re-radiated) by the oscillating MB can be calculated using \cite{56,57}:
\begin{linenomath*}
	\begin{equation}
	P_{sc}=\rho\frac{R}{r}(R\ddot{R}+2\dot{R}^2)
	\end{equation}
\end{linenomath*}
here $r$ is the distance from the MB center. The scattered pressure ($P_{sc}$) at 15 cm (approximate path length of the MBs in experiments) is calculated for 30 cycle pulses to match experimental conditions. The calculated $P_{sc}$ is then convolved with the one way transducer response accounting for attenuation effects in water ($0.000221 \frac{dB}{mmMHz^2}$ \cite{58}). Moreover, to better compare with experimental observations, the sampling frequency for simulations is chosen to be equal to the sampling frequency in experiments ($460 MHz$).  
\subsubsection{Investigation tools}
The results of the numerical simulations were visualized using a comprehensive bifurcation analysis method \cite{43}. In this method the bifurcation structure of the normalized MB oscillations ($\frac{R}{R_0}$) are plotted in tandem versus a control parameter using two different bifurcation methods (Poincaré section at each driving period and method of peaks). The bifurcation diagrams of the normalized MB oscillations ($R/R_0$) are calculated using both methods . When the two results are plotted alongside each other, it is easier to uncover more important details about the superharmonic (SuH) and ultraharmonic (UH) oscillations, as well as the SH and chaotic oscillations. This gives insight into the nonlinear behavior over a wide range of parameters, and enables the detection of SuH and UH oscillations alongside SH and chaotic oscillations \cite{43,POF,US}. This approach reveals the intricate details of the oscillations. In this paper the bifurcation diagrams of the normalized radial oscillations of the $Definity^{\textregistered}$ MBs were plotted versus the MB initial diameter for fixed frequencies of 25 MHz and for the range of the pressure values studied in the experiments.
\section{Results}
\subsection{Bifurcation structure}
In \ref{subsection:AA} we present the bifurcation structure of an uncoated C3F8 MB and a Definity MB with $2\mu m$ initial size as a function of frequency at $P_a=5 kPa$ for different values of initial surface tension $\sigma_0$. We show that the lipid buckling or rupture of the lipid shell enhances the unexpected generation of the nonlinear behavior.  Despite the reduced damping, the unocated MB of the same size does not exhibit nonlinear oscillations. In \ref{subsection:BB} we study the bifurcation structure of the uncoated and Definity MBs as a function of the initial size at different $\sigma_0$. The sonication frequency is fixed at 25 MHz (freqeuncy of the experiments) and $P_a=250 kPa$. We show that depending on the bubble size and $\sigma_0$, different nonlinear regimes can be enhanced.     
\subsubsection{Bifurcation of $\frac{R}{R_0}$ as a function of frequency for uncoated and coated MBs with different $\sigma(R_0)$} \label{subsection:AA}
\begin{figure}
	\begin{tabular}{c c}
		\includegraphics[scale=0.43]{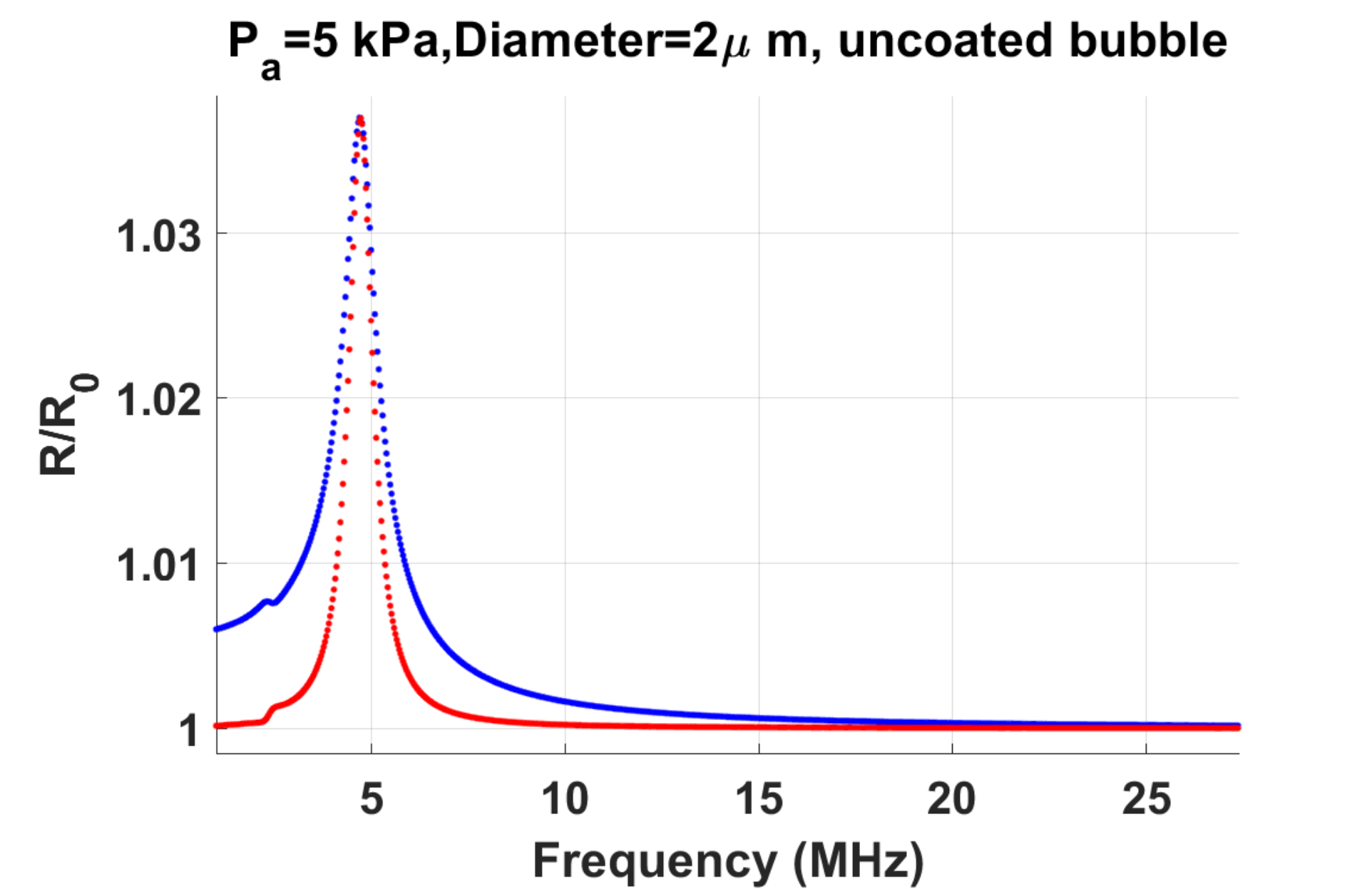} \hspace{-1cm}
		\includegraphics[scale=0.36]{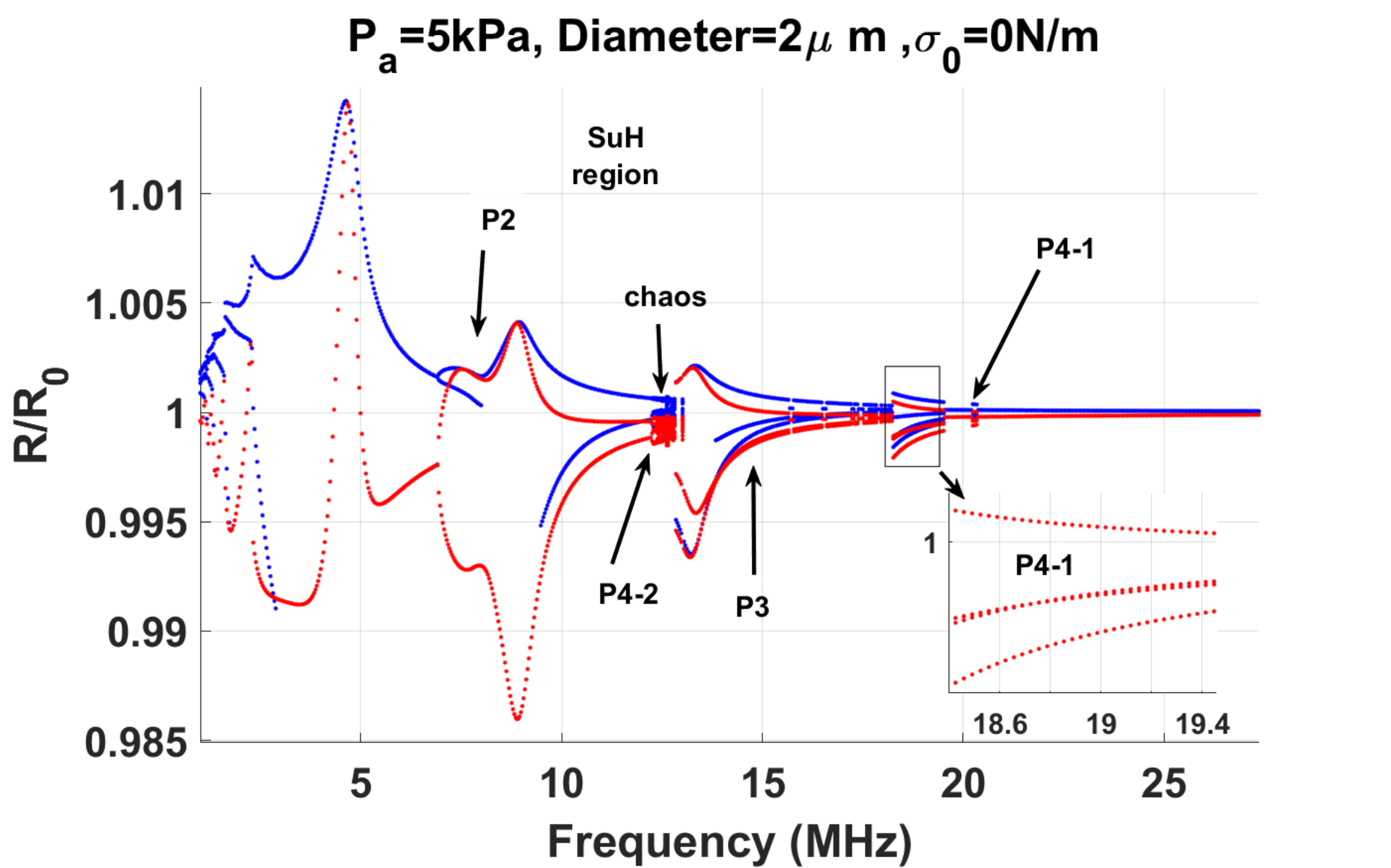}\\
		\hspace{-7cm}	(a) & \hspace{-5cm} (b)\\
		\includegraphics[scale=0.36]{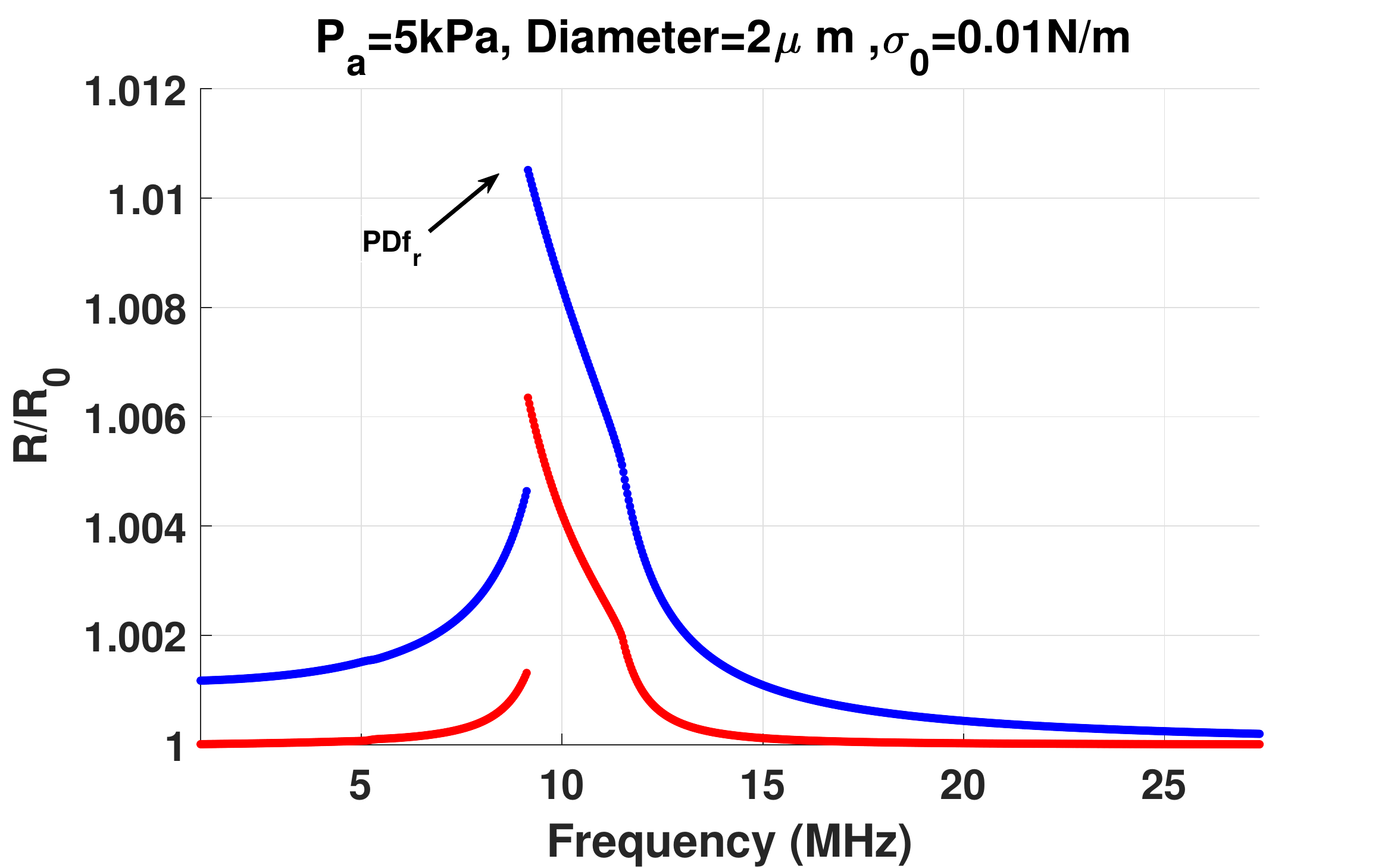} \hspace{-1cm}
		\includegraphics[scale=0.36]{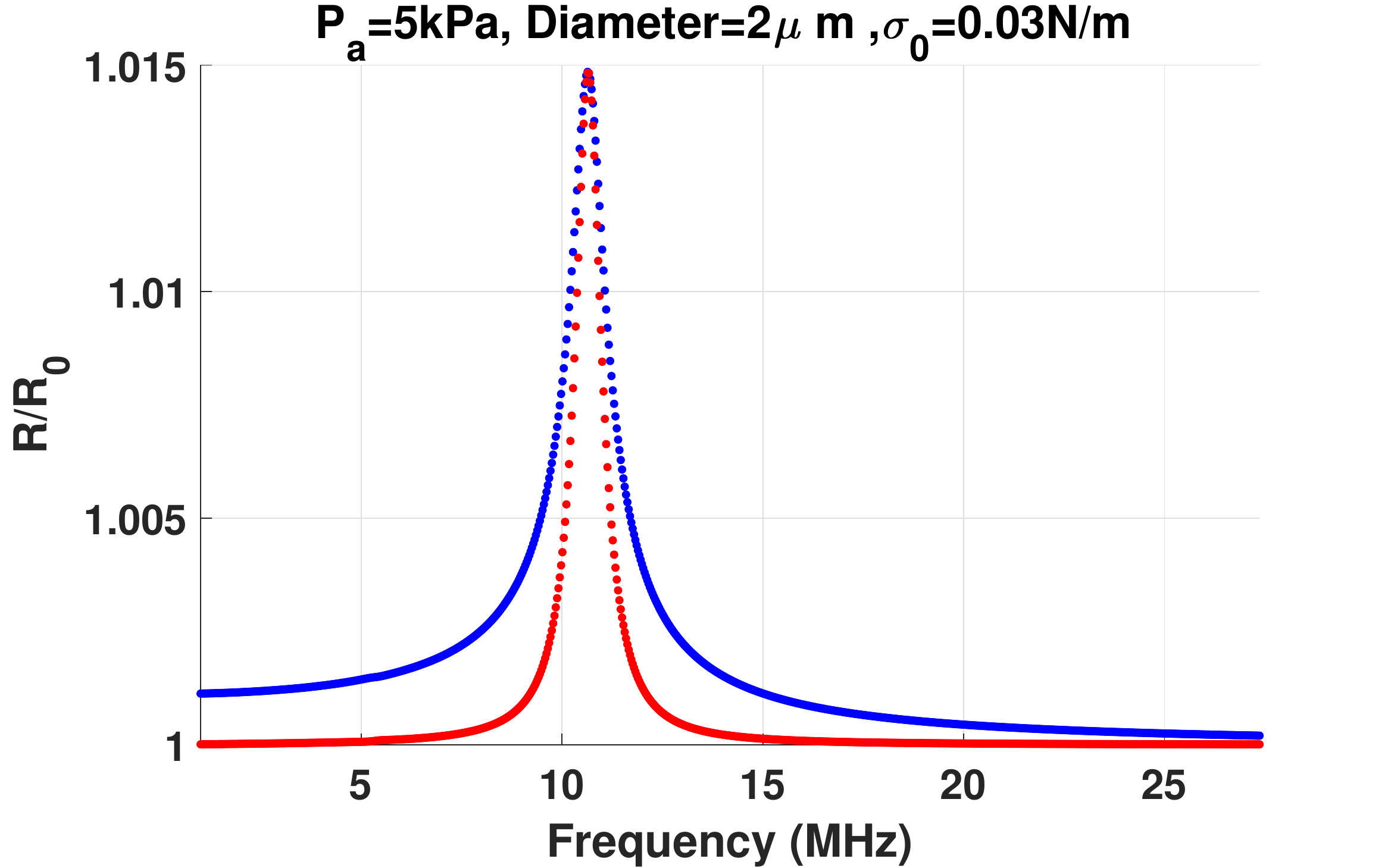}\\
		\hspace{-7cm}	(c) & \hspace{-5cm} (d)\\
		\includegraphics[scale=0.36]{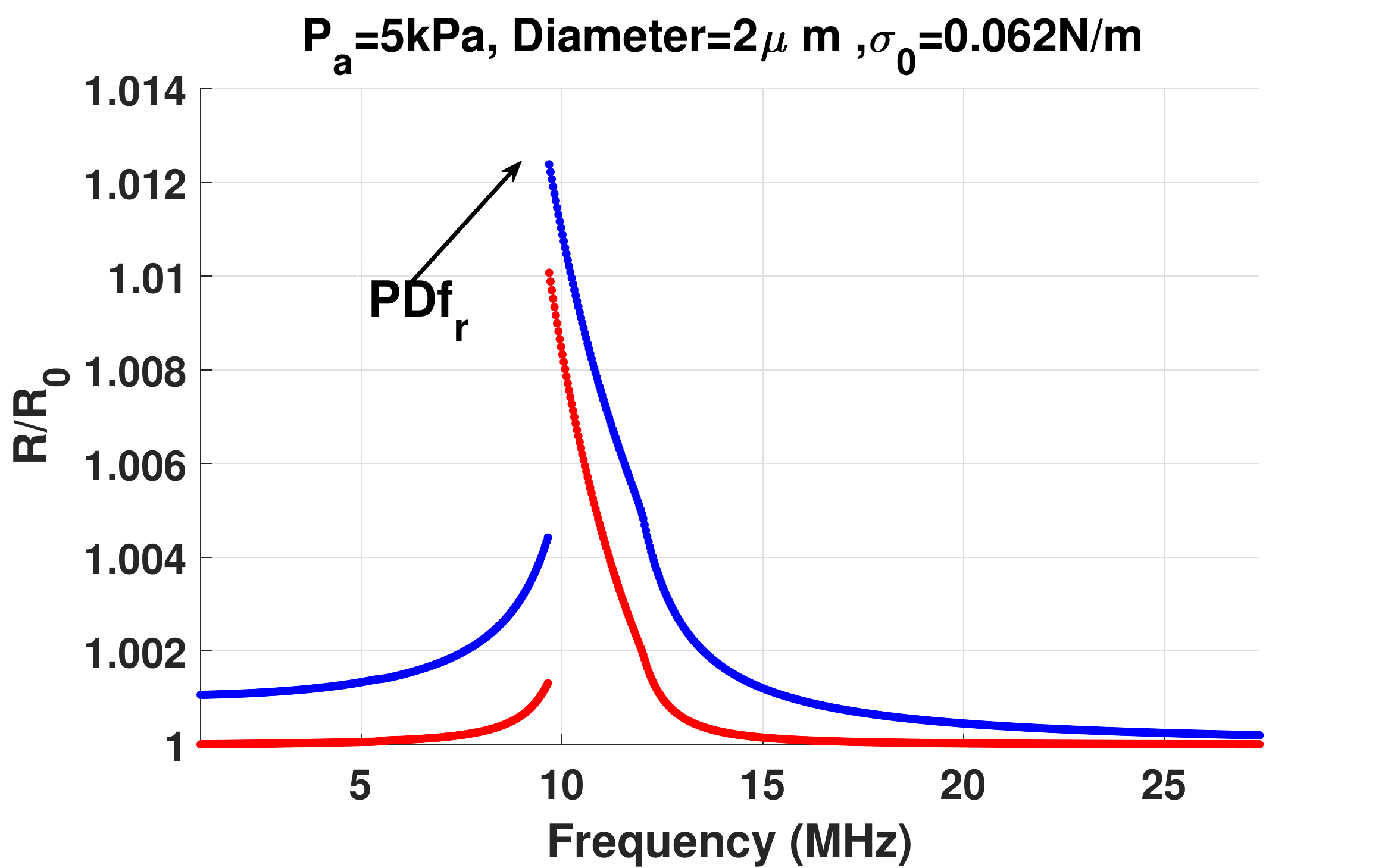} \hspace{-1cm}
		\includegraphics[scale=0.36]{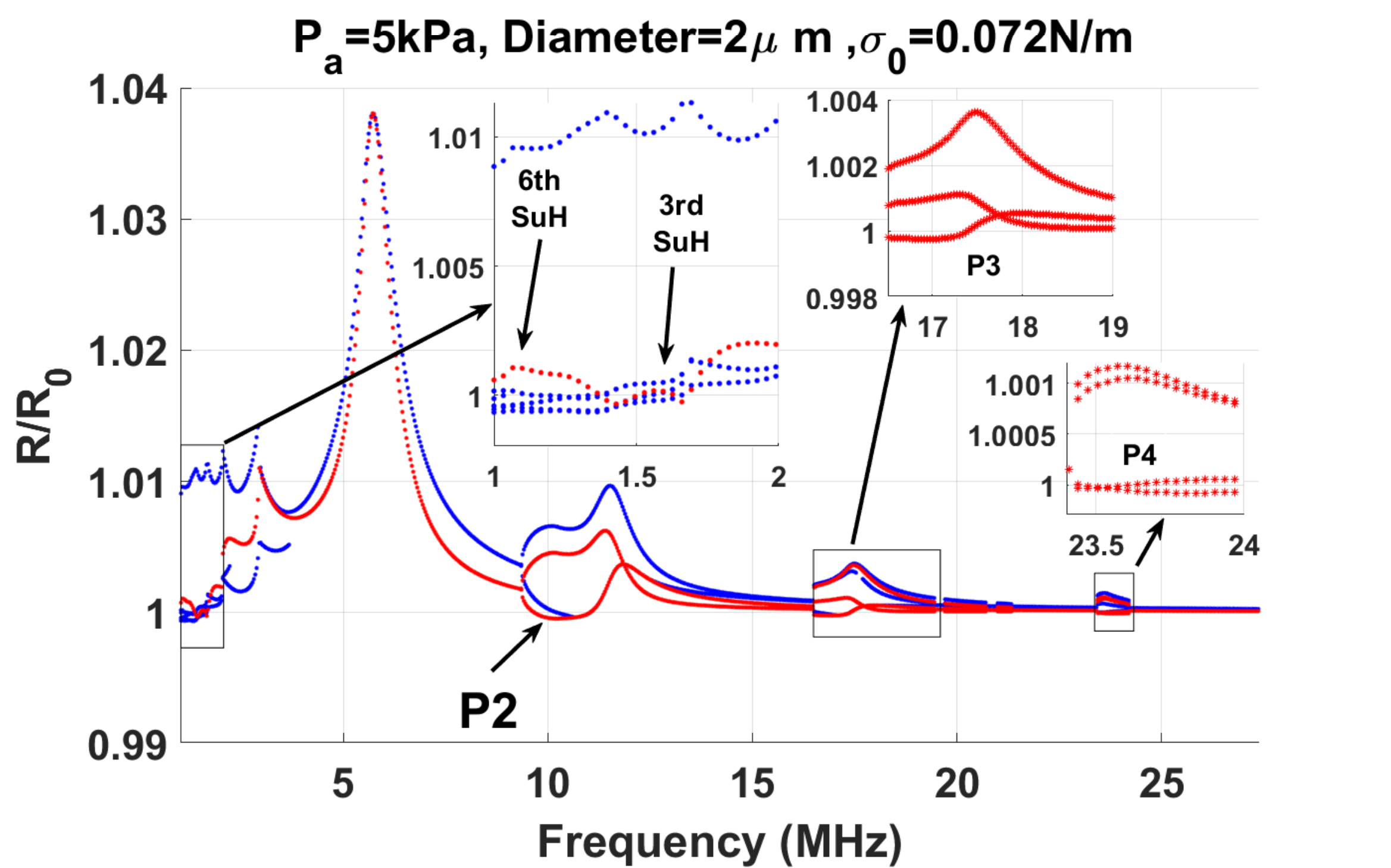} \\
		\hspace{-7cm}	(e) & \hspace{-5cm} (f)\\
	\end{tabular}
	\caption{The bifurcation structure (blue represent the maxima and the red represents the conventional method) of the $\frac{R}{R_0}$ of a $2 \mu m$ MB as a function of frequency for $P_a=5 kPa$ for the: a) uncoated MB and for the lipid MB with b) $\sigma(R_0)=0 N/m$, c) $\sigma(R_0)=0.01 N/m$, d) $\sigma(R_0)=0.03 N/m$, e) $\sigma(R_0)=0.062 N/m$ and f) $\sigma(R_0)=0.072 N/m$.}
\end{figure}
In order to visualize the dynamics of the MBs at low pressures, the bifurcation structure of a $2 \mu m$ MB is plotted as a function of frequency when $P_a=5 kPa$ (Fig. 2). The blue graph represents the results using the maxima method and the red graph represents the results using the conventional method (Poincaré section at each period). The uncoated MB (Fig. 2a) exhibits a P1 signal over 1-30 MHz with one maximum and resonant oscillations at $\approx 4.7 MHz$. Contrary to Fig. 2a, the lipid MB with $\sigma(R_0)=0$ exhibits significant non-linearity at 5 kPa (Fig. 2b). Higher order SuHs ($5th-2nd$ order) are seen for $f<\approx 2.88 MHz$. For example the 5th order SuH at 1MHz is a P1 signal with 5 maxima. P1 resonance occurs at $f\approxeq 4.63 MHz$ with P2 oscillations over a wide frequency range of $\approx$ $6.9MHz<f<12.22MHz$ with a small window of P4-2 and chaos. We call this P4-2 oscillations as it occurs when P2 oscillations undergo a period doubling (PD) to P4 \cite{23,44}. P3 occurs between $\approx$ $12.85MHz<f<18.2MHz$.  P4-1 occurs in the frequency range between $18.2MHz<f<19.5MHz$ (highlighted in subplot within Fig. 2b). We call this P4-1 regime as it occurs when P1 oscillations undergo a saddle node bifurcation to P4 oscillations \cite{26,29,30}.\\ The lipid MB with $\sigma(R_0)=0.01N/m$ (Fig. 2c) exhibits P1 oscillations with one maximum. The pressure dependent resonance frequency ($PDf_r$ \cite{42}) occurs at $f\approx 9.15 MHz$. The MB behavior is of P1 with one maximum for all the studied frequencies.\\ The lipid MB with $\sigma(R_0)=0.03 N/m$ (Fig. 2d) exhibits P1 behavior with 1 maximum and a resonance at $\approx 10.66MHz$. The behavior of the MB with $\sigma=0.062 N/m$ (Fig. 2e) is similar to $\sigma=0.01 N/m$ with $PDf_r\approxeq9.6 MHz$.\\ The MB with $\sigma(R_0)=0.072 N/m$ exhibits a similar behavior to the MB with $\sigma(R_0)=0 N/m$ demonstrating $6th-2nd$ order SuHs between the freqeuncy range of 1-3.7 MHz. Some of the SuH between $1MHz<f<2MHz$ are highlighted in a subplot within Fig. 2f). P2 occurs between $9.3MHz<f<16.5MHz$. P3 occurs between $16.5MHz<f<19 MHz$ (highlighted as a subplot inside Fig. 2f), between $19.7<f<20.7 MHz$ and between $20.3MHz<f<21.3 MHz$. P4-1 regime occurs between $23.4MHz<f<24 MHz$ (highlighted within a subplot in Fig. 2f).  The MB also demonstrates P1 resonance frequency at $\approx5.7 MHz$.\\Notably, despite the higher damping due to the coating, the coated MB undergoing shell rupture exhibits a greater oscillation amplitude (Figs. 2a and 2f). 
\begin{figure}
	\begin{tabular}{c c}
		\includegraphics[scale=0.36]{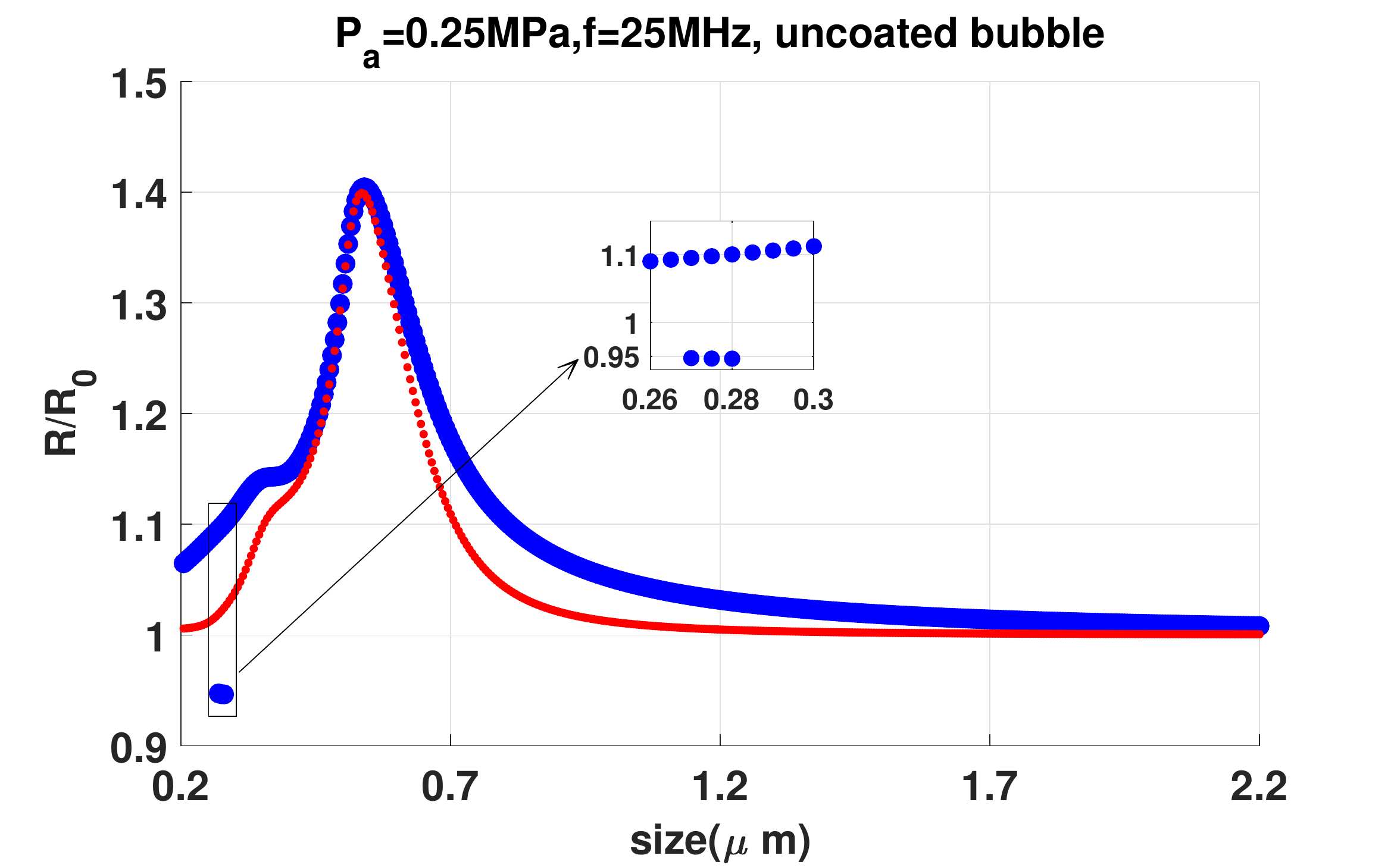} \hspace{-1cm}
		\includegraphics[scale=0.36]{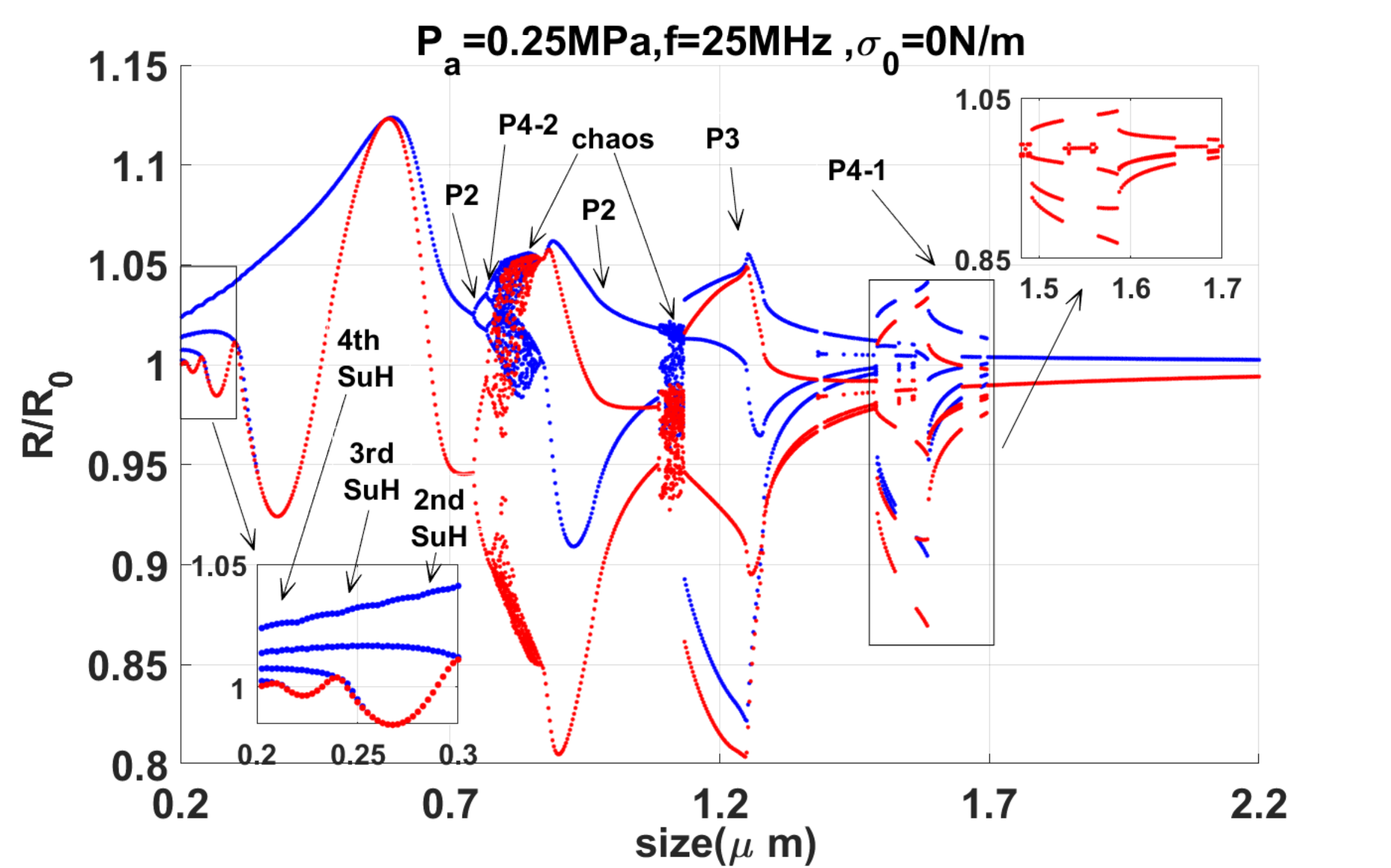}\\
		\hspace{-7cm}	(a) & \hspace{-5cm} (b)\\
		\includegraphics[scale=0.36]{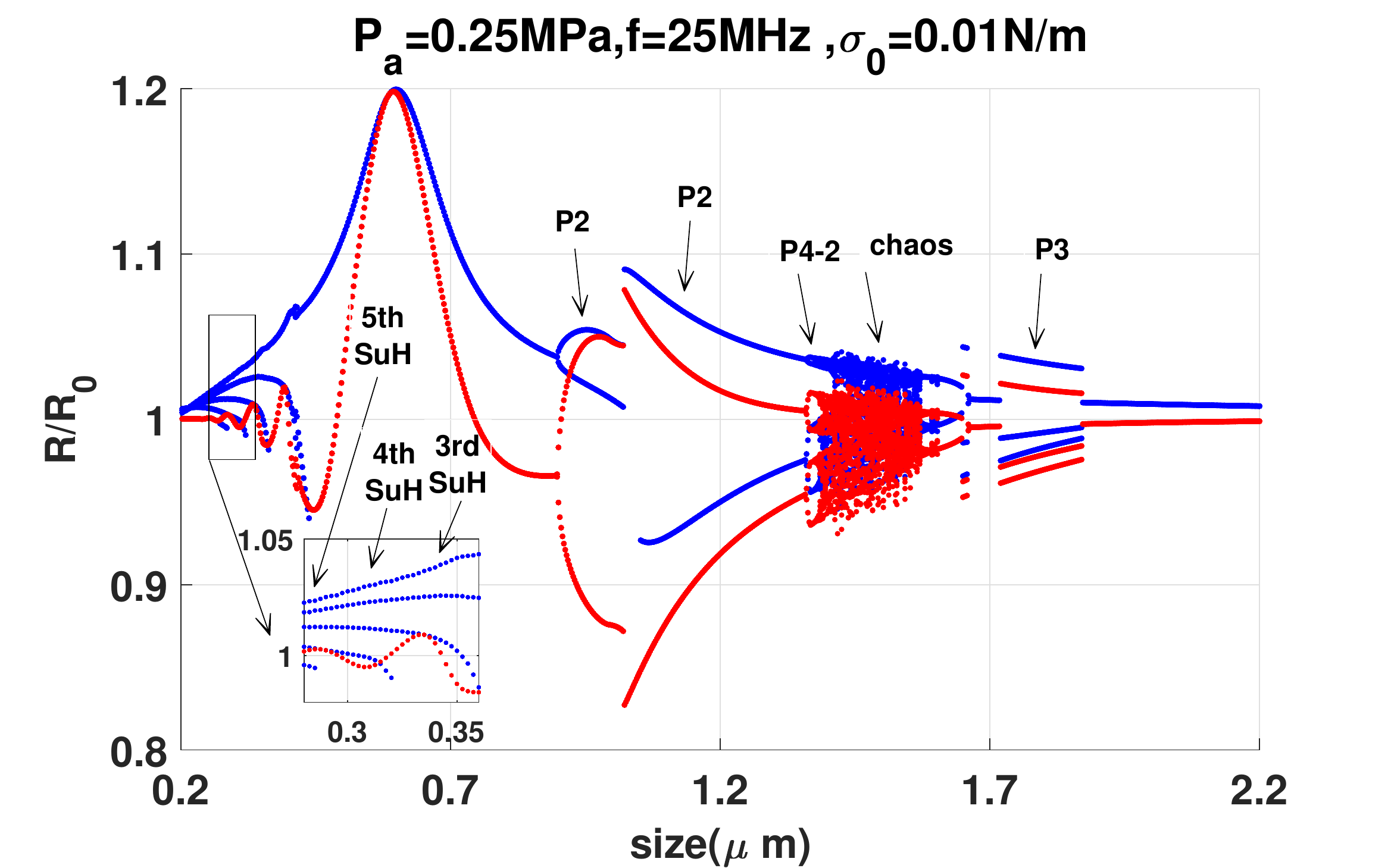} \hspace{-1cm}
		\includegraphics[scale=0.36]{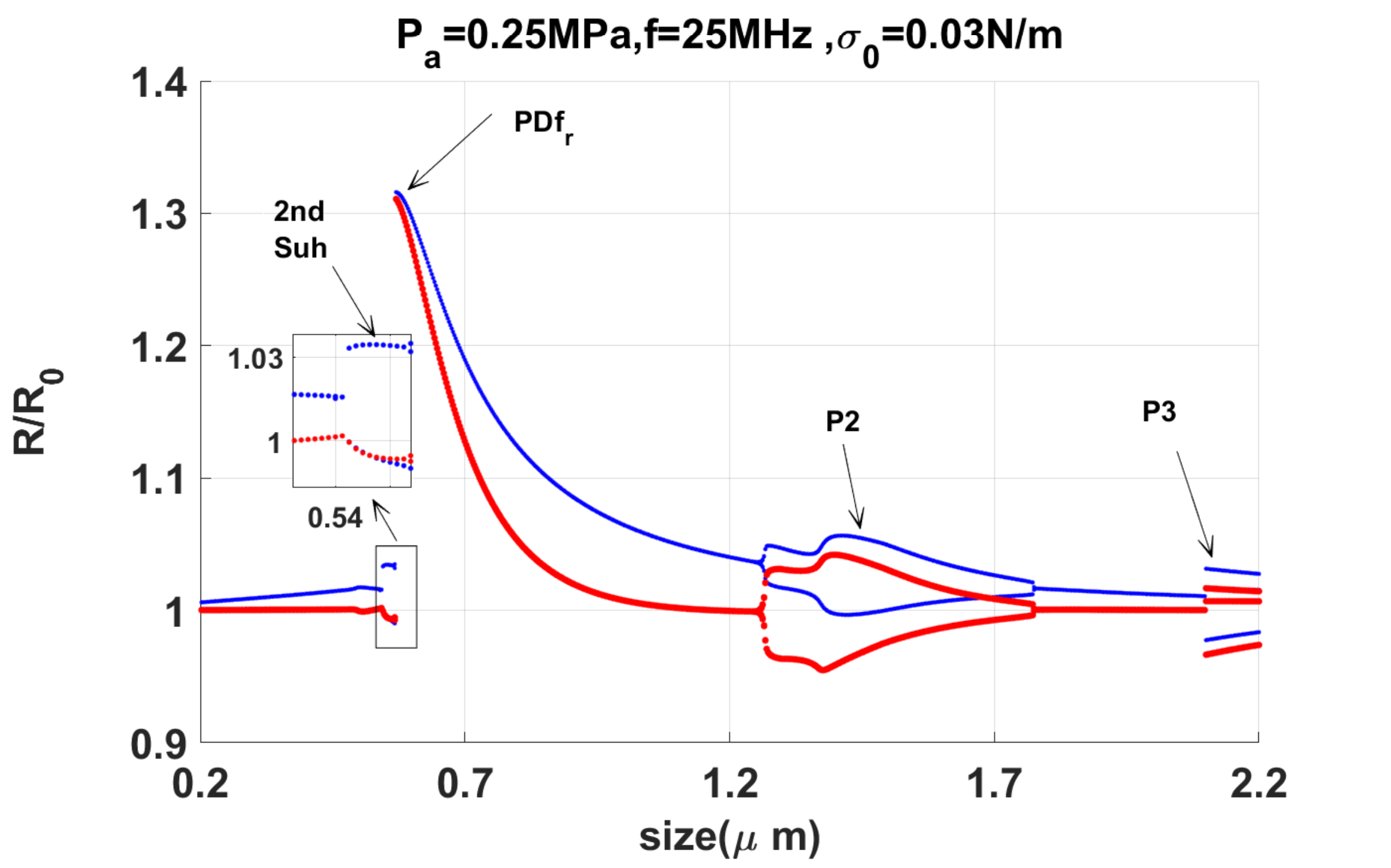}\\
		\hspace{-7cm}	(c) & \hspace{-5cm} (d)\\
		\includegraphics[scale=0.36]{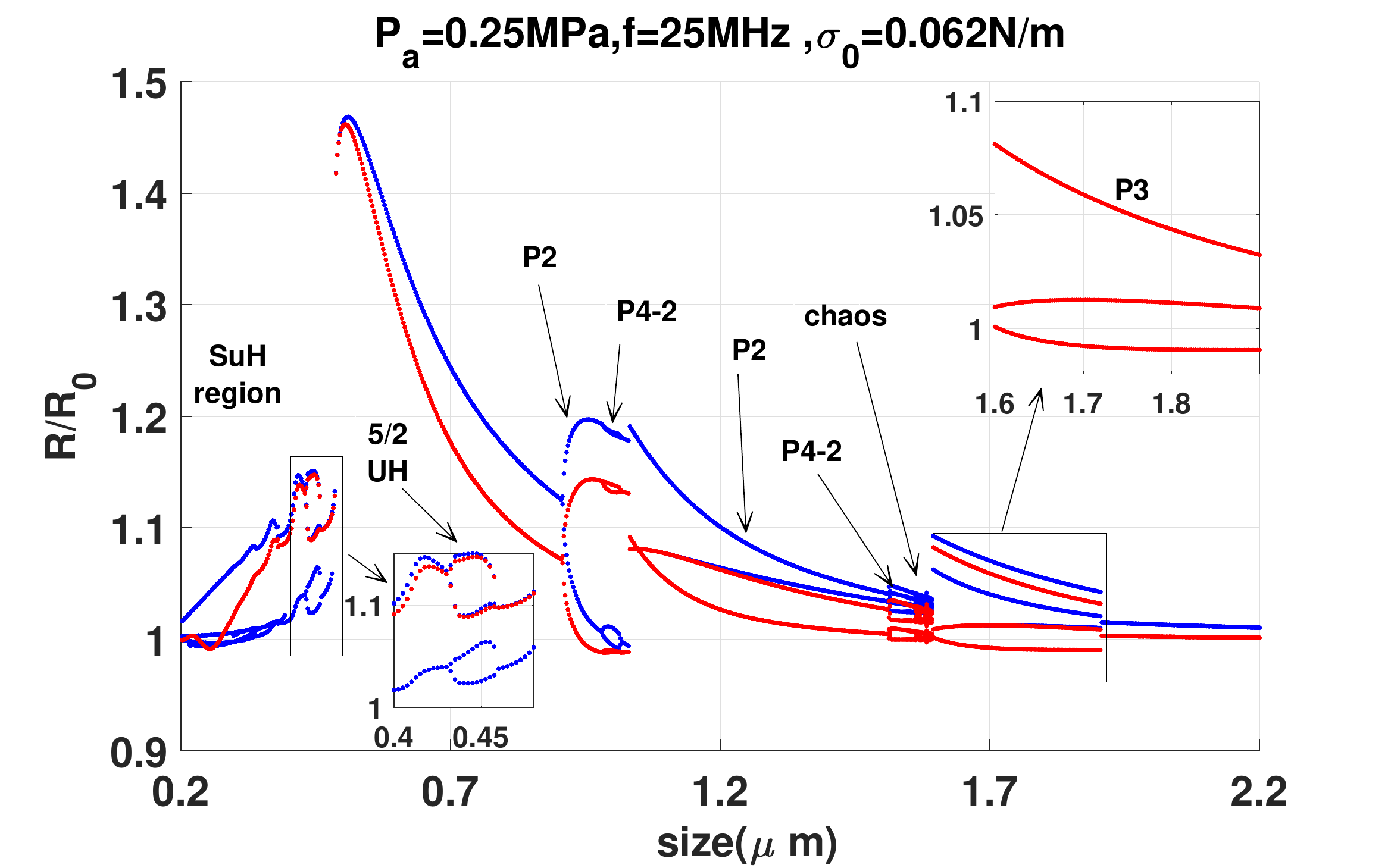} \hspace{-1cm}
		\includegraphics[scale=0.36]{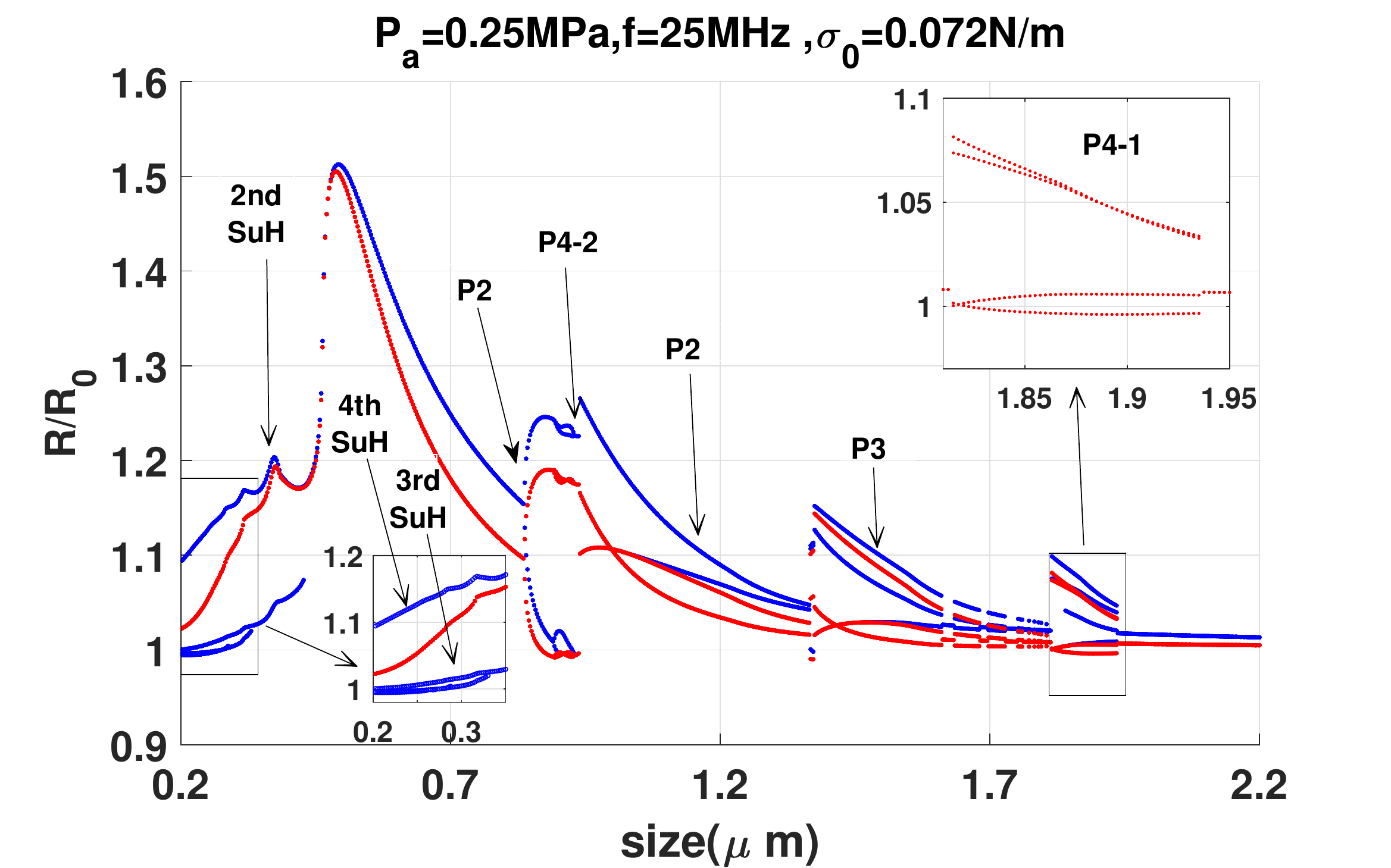} \\
		\hspace{-7cm}	(e) & \hspace{-5cm} (f)\\
	\end{tabular}
	\caption{The bifurcation structure of the $\frac{R}{R_0}$ (blue represent the maxima and the red represents the conventional method) as a function of size (MB diameter) at $P_a=250 kPa$ and $f=25 MHz$ for the: a) uncoated MB and for the lipid MBs with b)$\sigma(R_0)=0 N/m$, c)$\sigma(R_0)=0.01 N/m$, d)$\sigma(R_0)=0.03 N/m$, e) $\sigma(R_0)=0.062 N/m$ and f) $\sigma(R_0)=0.072 N/m$.}
\end{figure}
\subsubsection{Bifurcation $\frac{R}{R_0}$ as a function of size (initial diameter) for uncoated and coated MBs with different $\sigma(R_0)$}
\label{subsection:BB}
In order to investigate the nonlinear behavior of the commercially available $Definity^{\textregistered}$ MBs for the experimental exposure conditions,  the bifurcation structure of the $\frac{R}{R_0}$ is studied as a function of MB size when $P_a=250 kPa$ and $f=25MHz$. This is because of the polydisperse nature of the $Definity^{\textregistered}$ MBs \cite{54}, and since in the experiments we are limiting our analysis to the transducer focal zone with small variations in pressure and the fixed sonication frequency. The size distribution in the simulations replicates the distribution of the native $Definity^{\textregistered}$ \cite{54} Thus, the $\frac{R}{R_0}$ plot versus MB size will provide insight relevant to the experimental conditions in this study.\\
Fig. 3a, shows the bifurcation structure of an uncoated MB as a function of size. MB with sizes between $0.27 \mu m$-$0.28 \mu m$ exhibit 2nd order SuH (P1 oscillation with 2 maxima as highlighted in a subplot) and MBs with sizes $0.54 \mu m$ are resonant. Fig. 3a shows that at $f=25 MHz$ and $P_a=250 kPa$ the uncoated MB cannot produce SHs.\\ Fig. 3b, shows the bifurcation structure of the $Definity^@$ MBs with $\sigma(R_0)=0 N/m$. In stark contrast to the uncoated MB (Fig. 3a), an abundance of nonlinear behavior is observed. This includes 4th, 3rd and 2nd order SuHs for MB sizes smaller than $0.345 \mu m$ (some are highlighted in a subplot within Fig. 3b), P1 resonance for $\approx 0.59 \mu m$ MBs , P2, P4-2 and chaotic behavior for MB sizes of $\approx$ $0.74 \mu m<2R_0<1.13\mu m$, P3 oscillations for   $\approx$ $1.13 \mu m<2R_0<1.49\mu m$, and intermittent P4-1 oscillations for  $\approx$ $1.49 \mu m<2R_0<1.7\mu m$ (P4-1 is highlighted in a subplot within Fig. 3b).\\ $Definity^{\textregistered}$ MBs with $\sigma(R_0)=0.01 N/m$ (Fig. 3c), exhibit enhanced nonlinear behavior including $5th-2nd$ order SuHs (highlighted in a subplot), P2, P4-2, P3 and chaos. Fig. 3d represents the behavior of MBs with $\sigma(R_0)=0.03 N/m$. 2nd order SuH (highlighted in a subplot), $PDf_r$, P2 and P3 oscillations are observed.\\ Fig. 3e-f represent the MBs with initial surface tension close to that of water and thus with a higher tendency for rupture and expansion dominated behavior. For $\sigma(R_0)=0.062 N/m$ and for MB sizes $0.2\mu m<2R_0<0.5 \mu m$,  3rd and 2nd order SuH and 5/2 UH regimes are observed. 5/2 UH is a P2 with 4 maxima and is highlighted in a subplot within Fig. 3e. $PDf_r$, P2, P4-2 and P3 (highlighted in a subplot) oscillations are observed for MB sizes $2R_0>0.5 \mu m$. When $\sigma(R_0)=0.072 N/m$ (Fig. 3f), in addition to the nonlinear behavior we observe in Fig. 3e, we observe a 4th order SuH regime (highlighted in a subplot as a P1 with 4 maxima) and P4-1 and the absence of 5/2 UHs.\\ Results indicate that the nonlinear behavior of the MBs is highly sensitive to the initial surface tension as well as the MB size. The closer the surface tension to 0 or that of water ($\sigma_{water}=0.072 N/m$), the greater is the tendency of the MB to exhibit nonlinear behavior. Notably, P4-1 oscillations \textit{were only observed} when $\sigma(R_0)=0$ and $0.072 N/m$ (Fig. 3).\\ The MBs with $\sigma(R_0)=0.062$ (Fig. 3e) and $\sigma(R_0)=0.072$ (Fig. 3f), exhibit higher oscillation amplitude compared to uncoated MBs of the same size.\\ In order to better visualize the effect of the $\sigma(R_0)$ on the MB behavior, the bifurcation structure of the $\frac{R}{R_0}$ of the MB  is plotted as a function of  $\sigma(R_0)$ for two different MB sizes in Fig. 4. The bifurcation structure of a MB  with an initial diameter of $0.92 \mu m$ is depicted in Fig. 4a. The nonlinear behavior occurs only for the two extreme ends of the $\sigma(R_0)$. P2 occurs for $\sigma(R_0)<0.011 N/m$ and $\sigma(R_0)>0.061 N/m$ with P4-2 happening for $0.0032 N/m<\sigma(R_0)<0.048 N/m$  and $\sigma(R_0)>0.069 N/m$. For a MB  with initial diameter of $1.89 \mu m$, the same general behavior is observed. For initial surface tension values between $0.0127<\sigma(R_0)<0.057 N/m$ we observe P1 behavior with 1 maximum. As we approach to the lower and higher $\sigma(R_0)$, nonlinear behavior manifests itself in the bifurcation diagrams. P4-1 oscillations occurs for $0.0035 N/m<\sigma(R_0)<0.0041 N/m$ and $\sigma(R_0)>0.07 N/m$. P3 occurs for $0.009 N/m<\sigma(R_0)<0.012 N/m$ and $0.058 N/m<\sigma(R_0)<0.069 N/m$.
\begin{figure*}
	\begin{center}
		\includegraphics[scale=0.4]{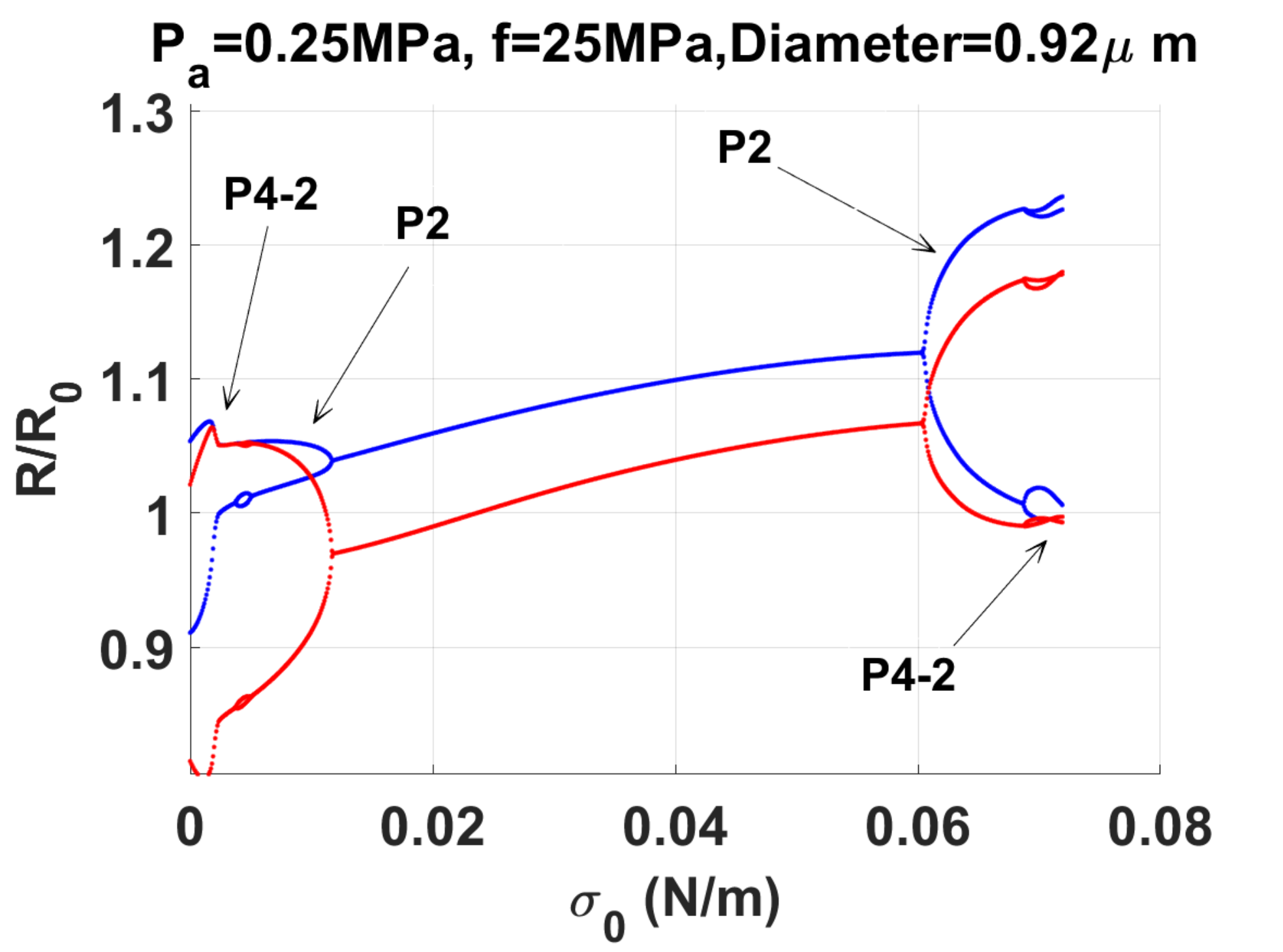}\includegraphics[scale=0.4]{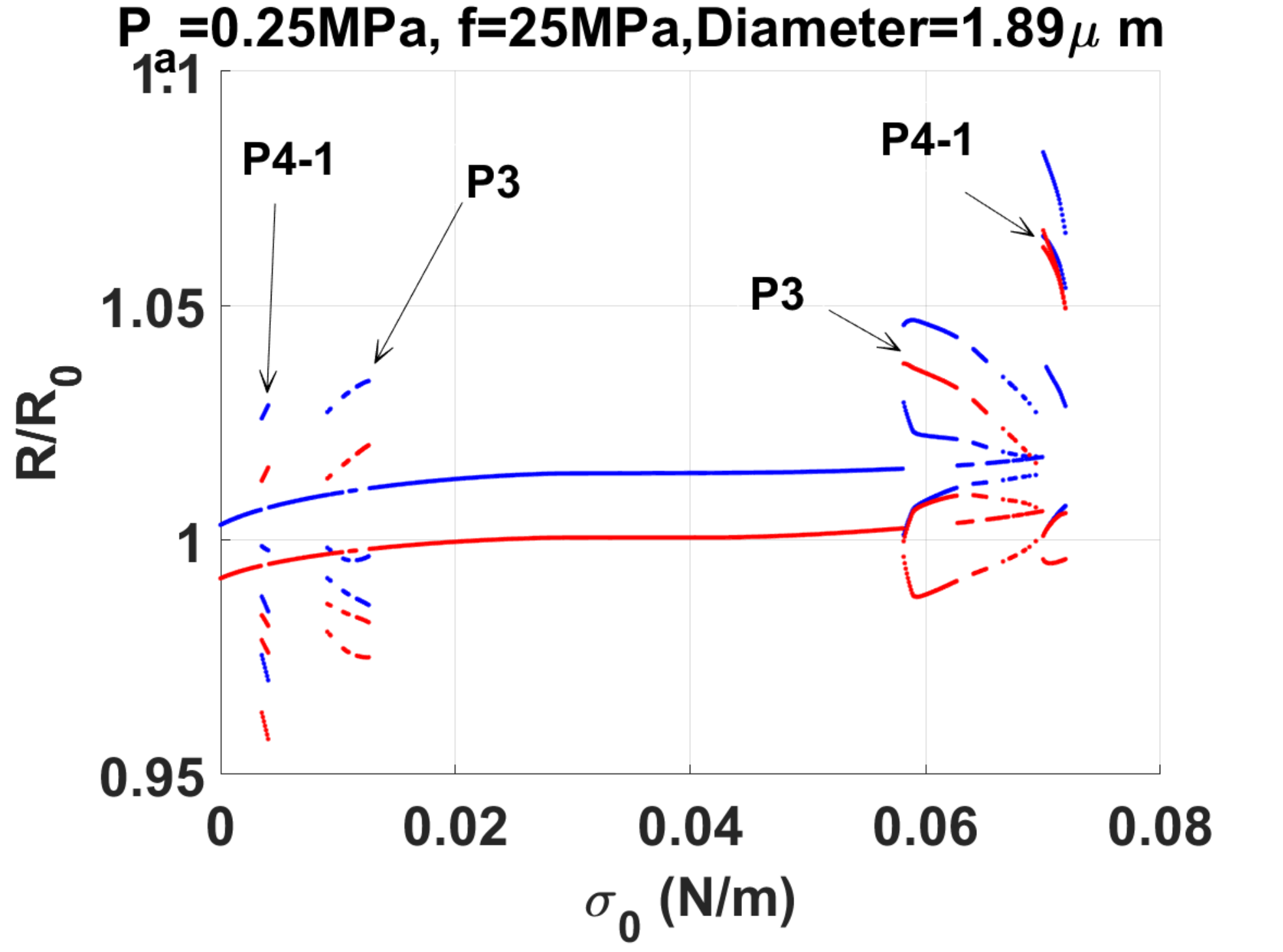}\\
		\hspace{2cm}	(a) \hspace{7cm}(b)
		\caption{The bifurcation structure of $\frac{R}{R_0}$ as a function of $\sigma(R_0)$ at $f=25 MHz$ and $P_a=250 kPa$ for a MB size of: a) 0.92 $\mu m$ $\&$ b) 1.89 $\mu m$ }
	\end{center}
\end{figure*}

\subsubsection{Experiments}
\begin{figure*}
	\begin{center}
		\includegraphics[scale=0.3]{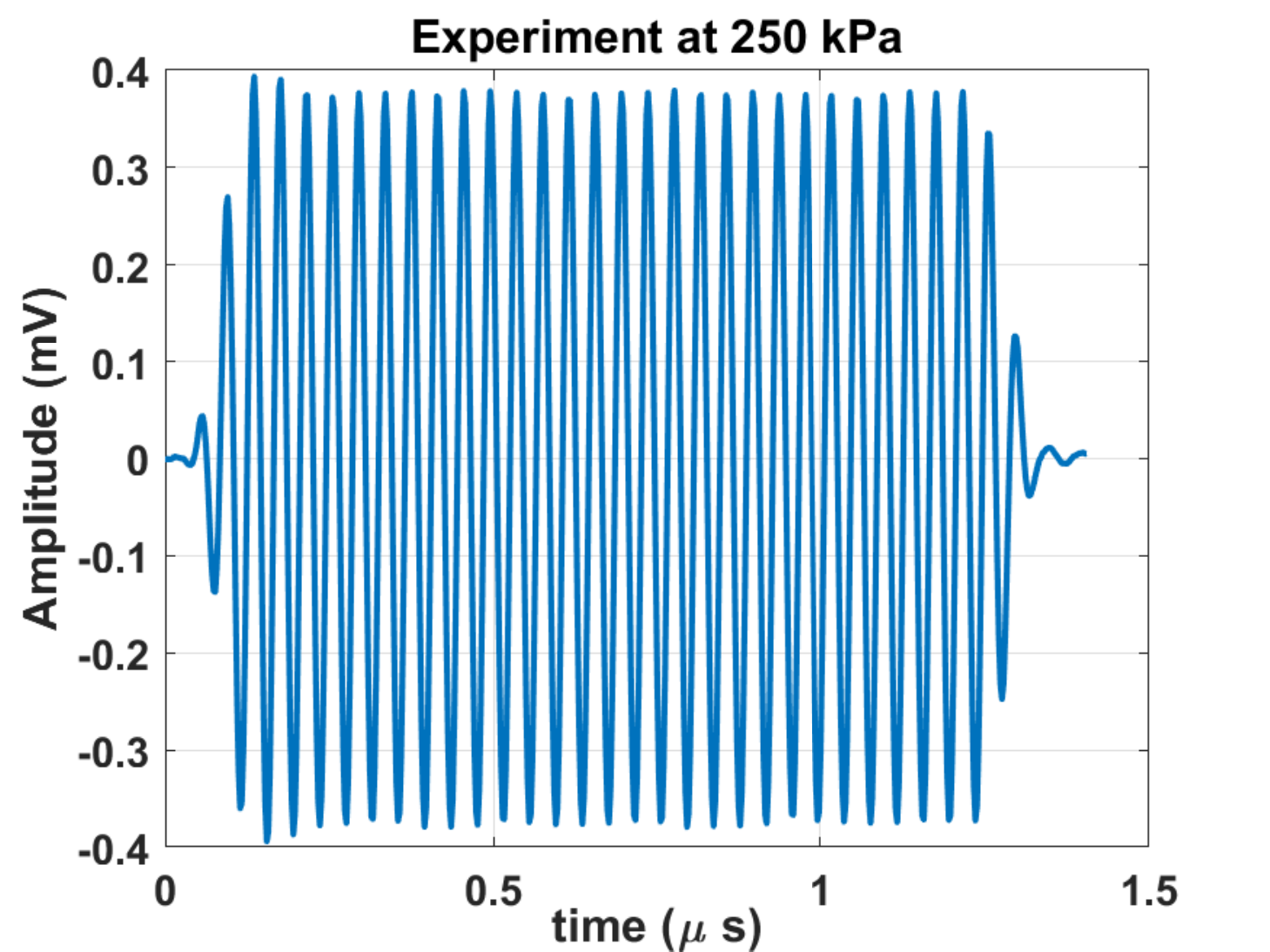}\includegraphics[scale=0.3]{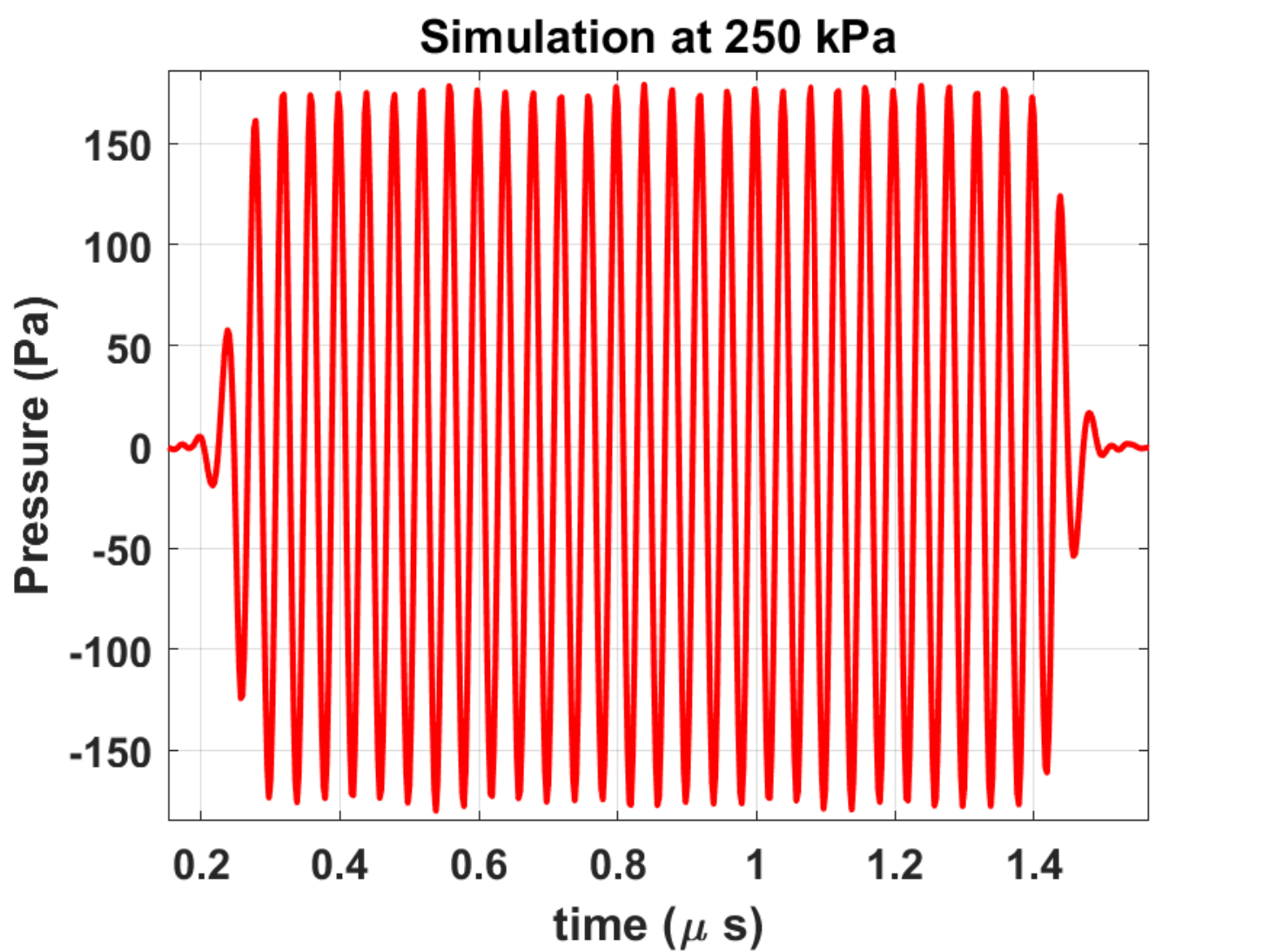}	\includegraphics[scale=0.3]{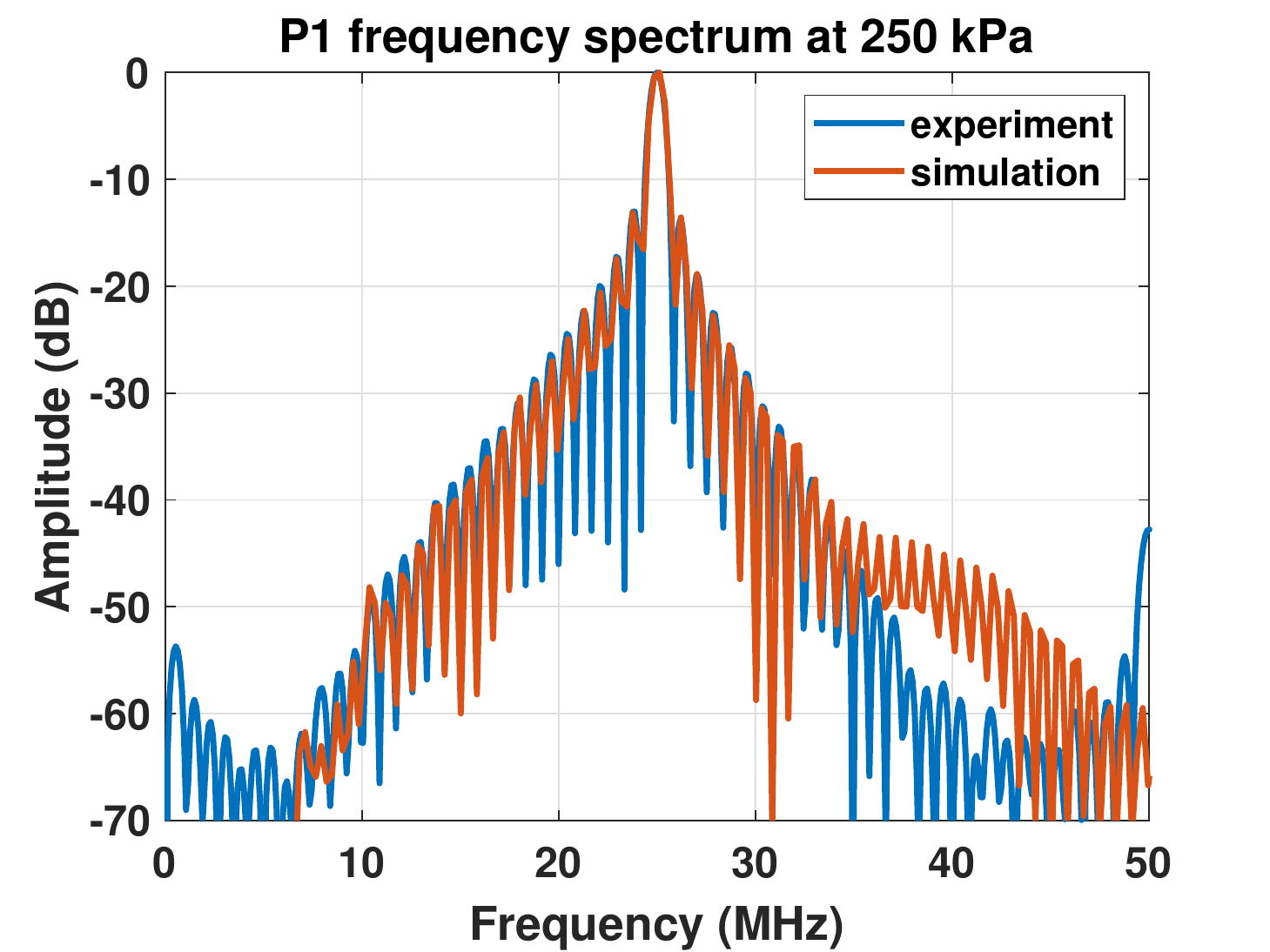}\\
		(a) \hspace{4cm}(b)\hspace{4cm}(c)\\
		\includegraphics[scale=0.3]{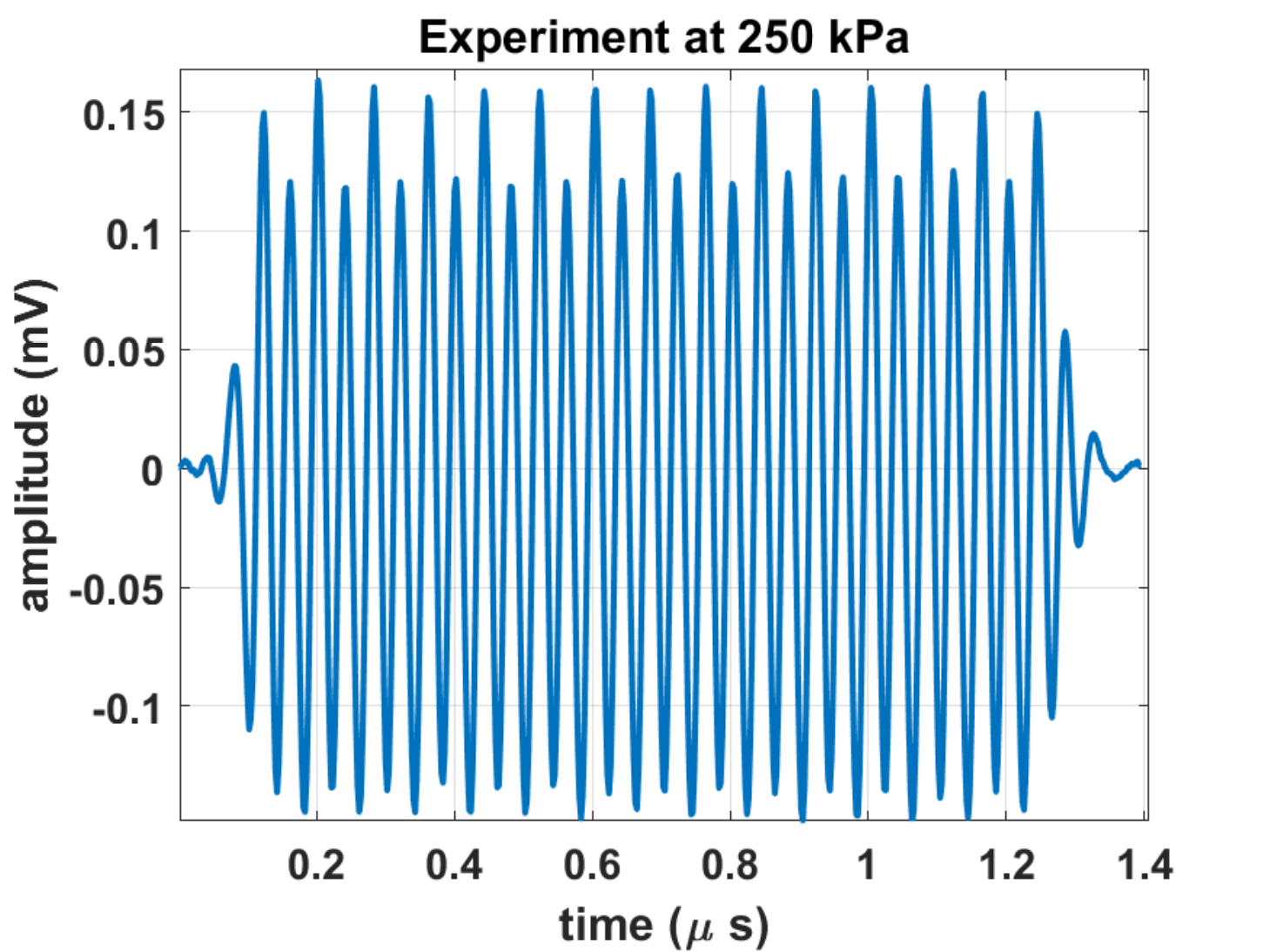}\includegraphics[scale=0.3]{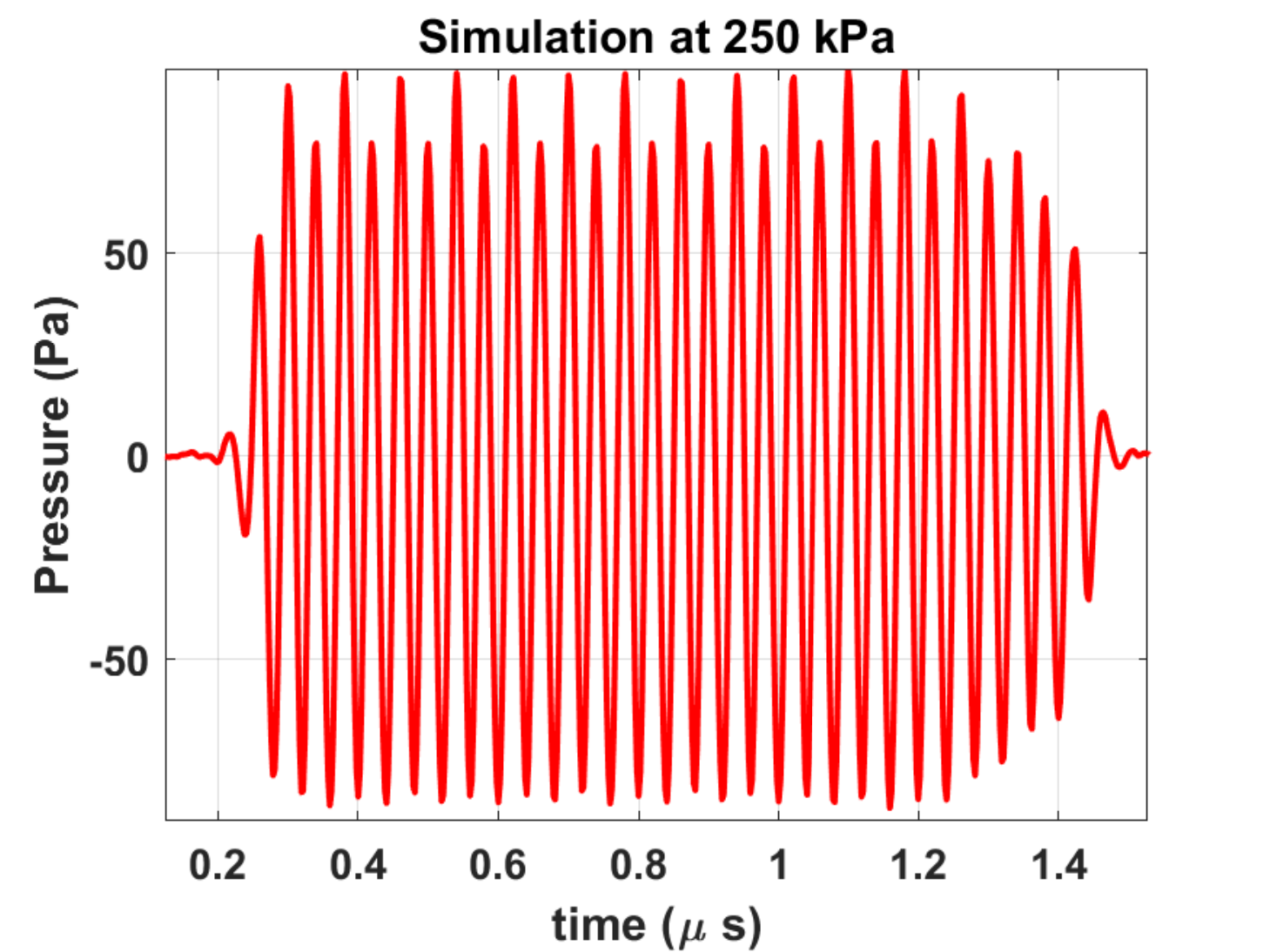}	\includegraphics[scale=0.3]{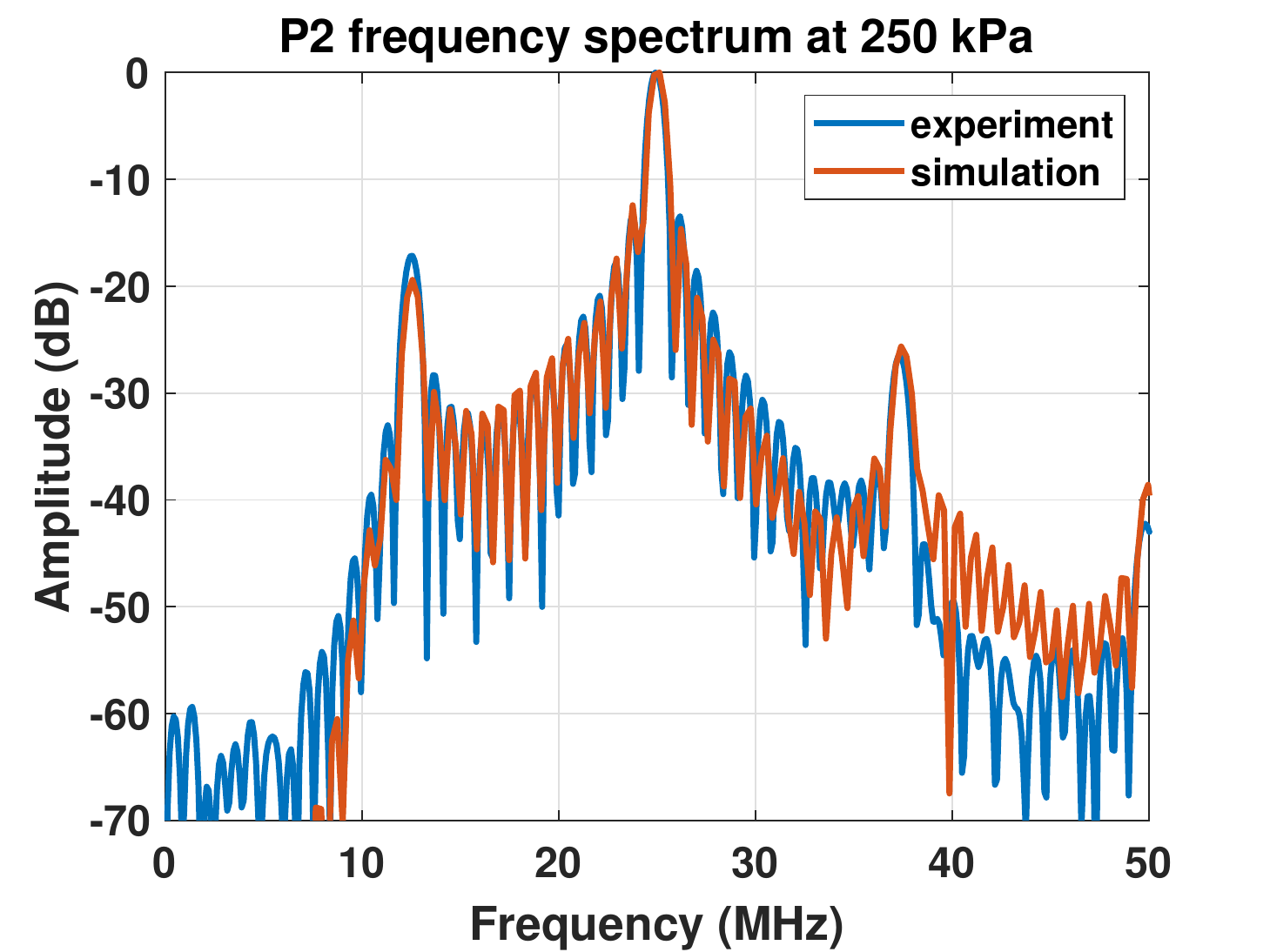}\\
		(d) \hspace{4cm}(e)\hspace{4cm}(f)\\
		\includegraphics[scale=0.3]{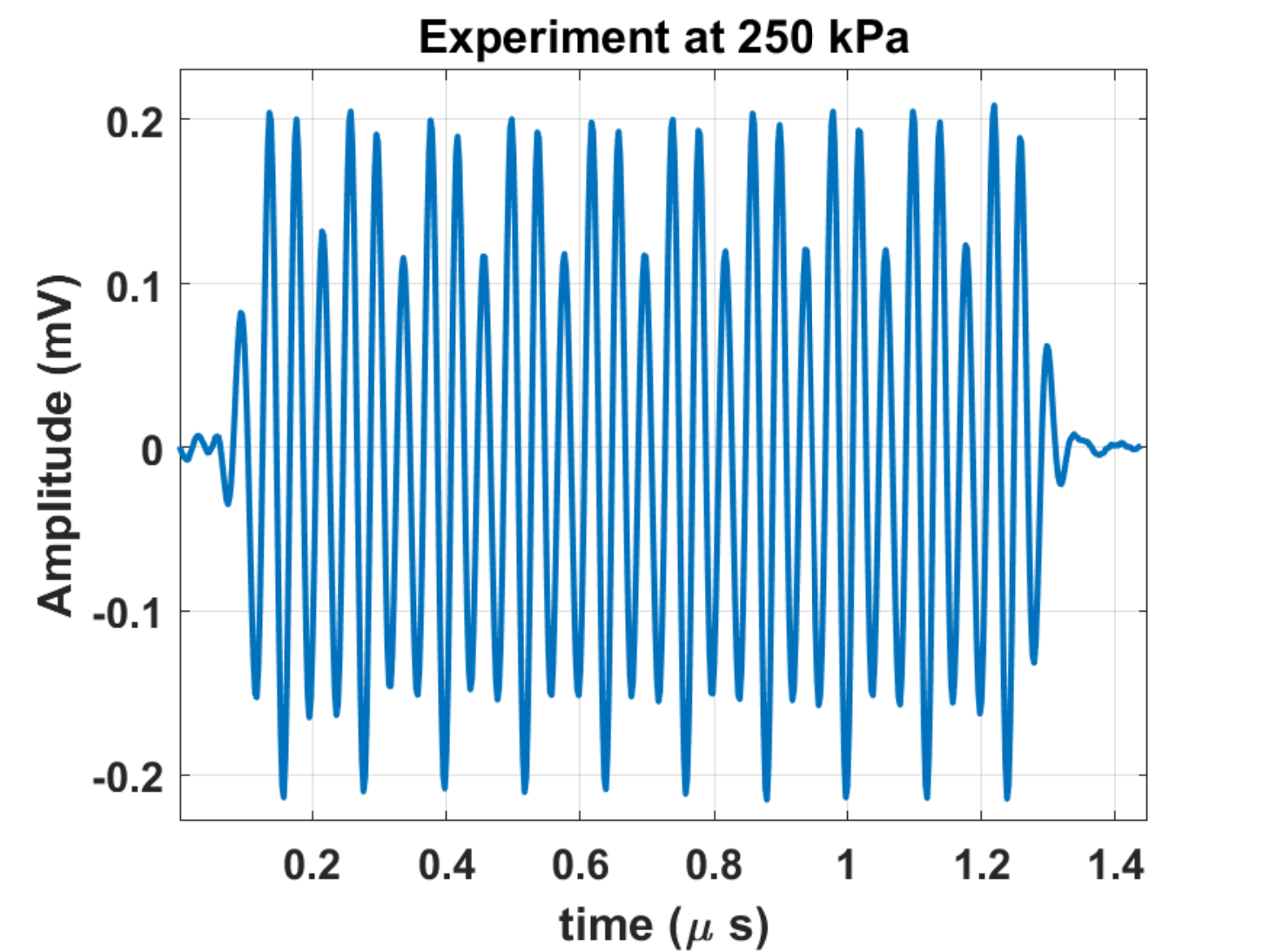}\includegraphics[scale=0.3]{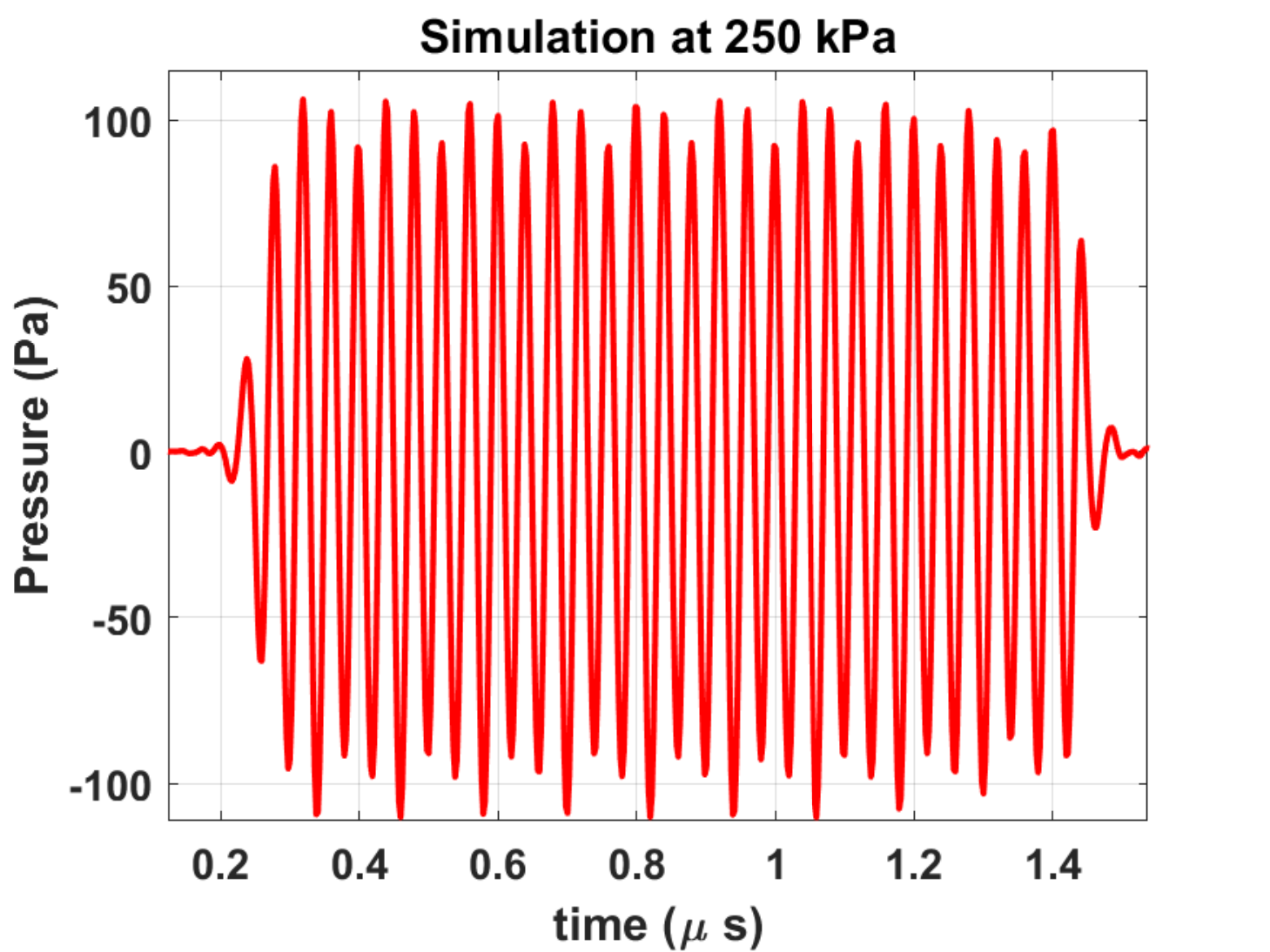}	\includegraphics[scale=0.3]{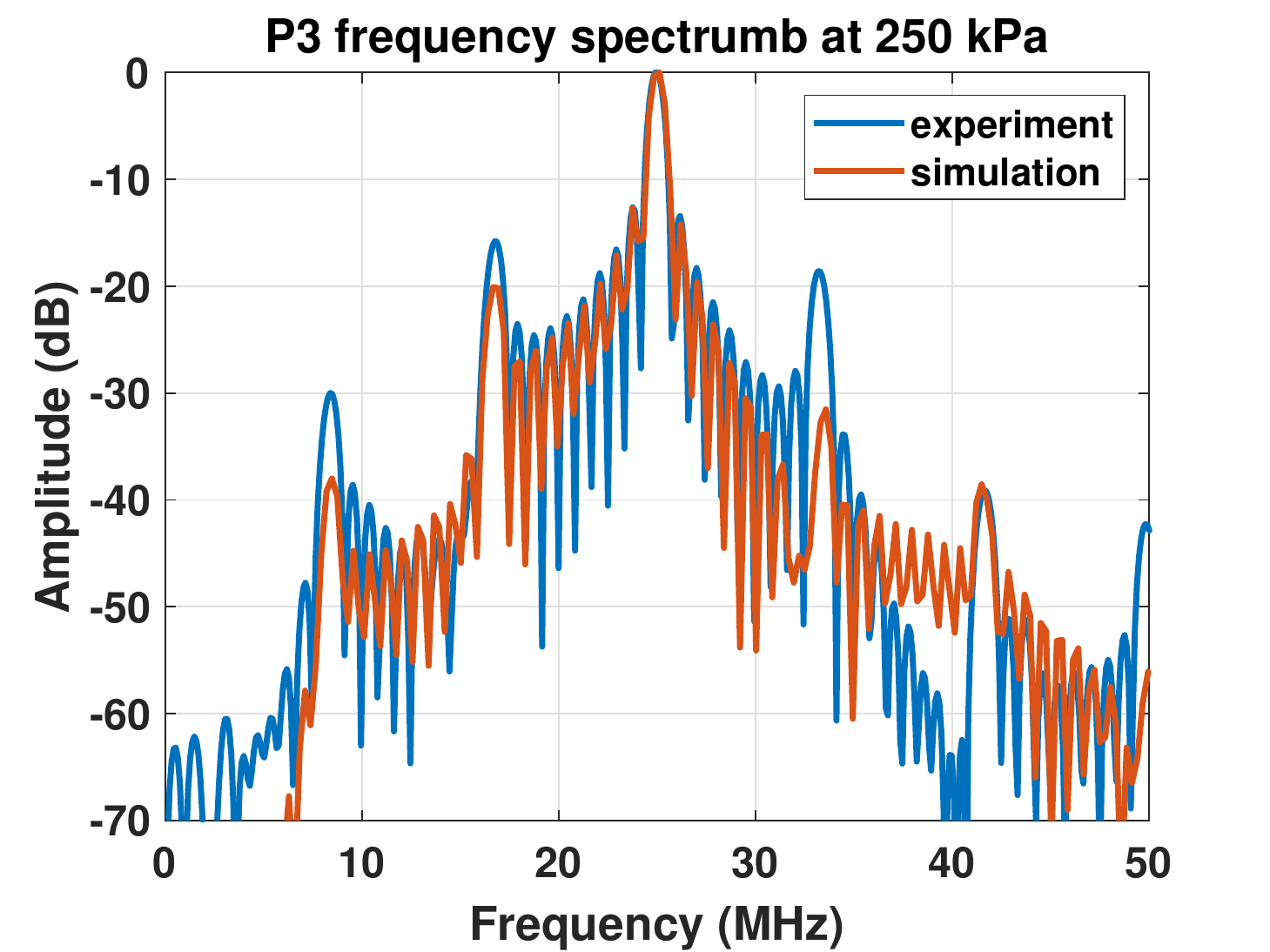}\\
		(g) \hspace{4cm}(h)\hspace{4cm}(i)\\
		\includegraphics[scale=0.3]{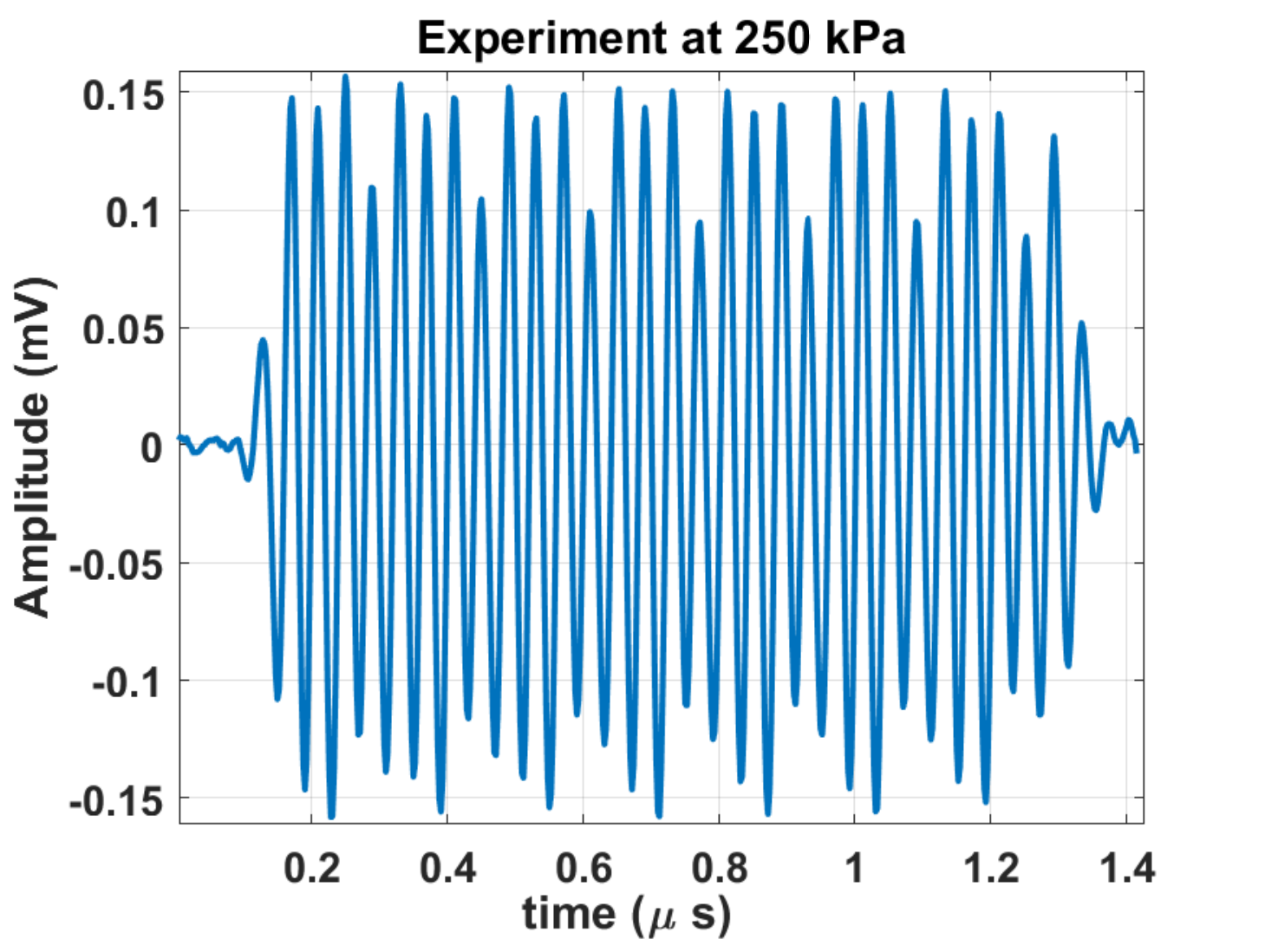}\includegraphics[scale=0.3]{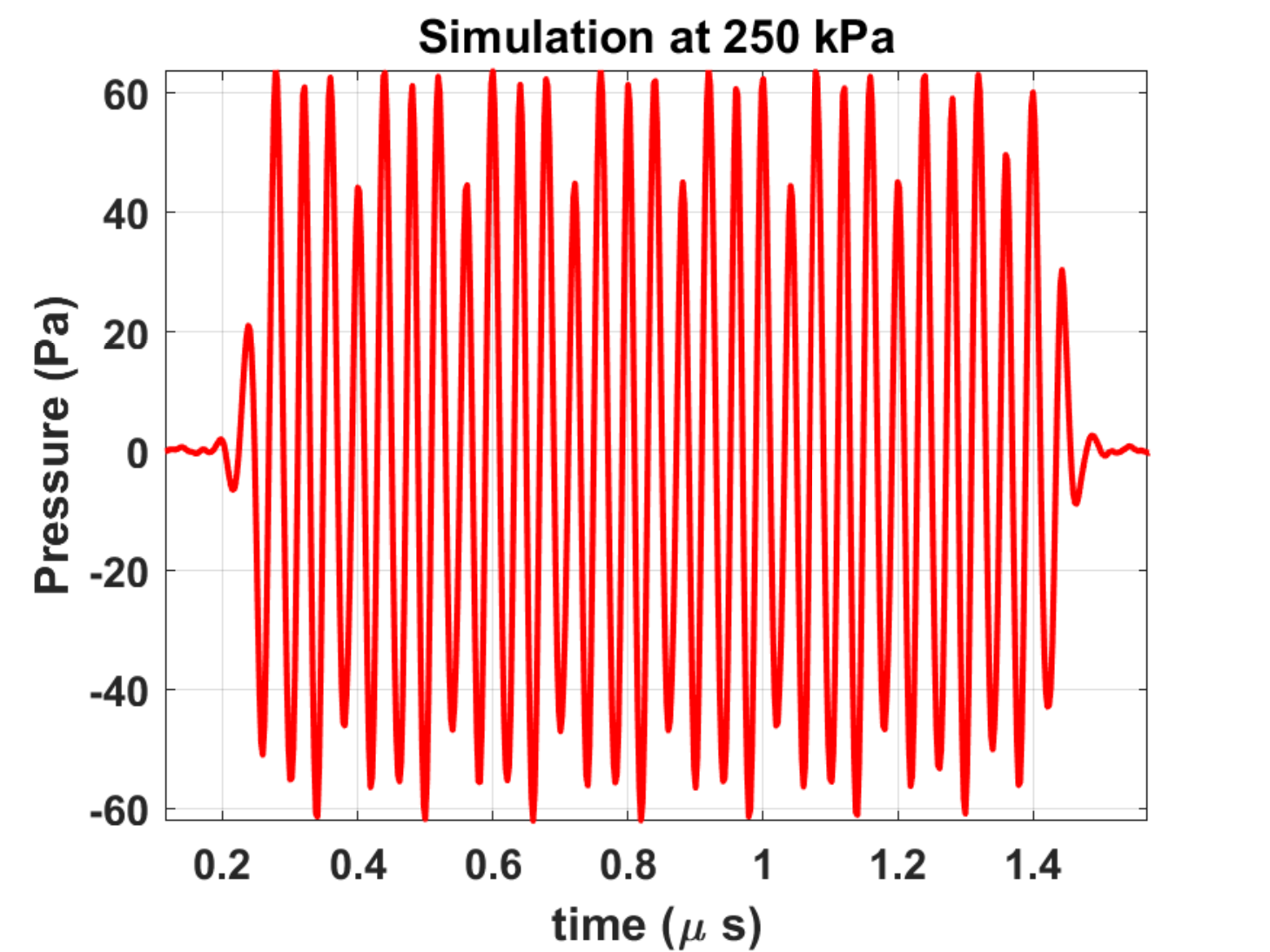}	\includegraphics[scale=0.3]{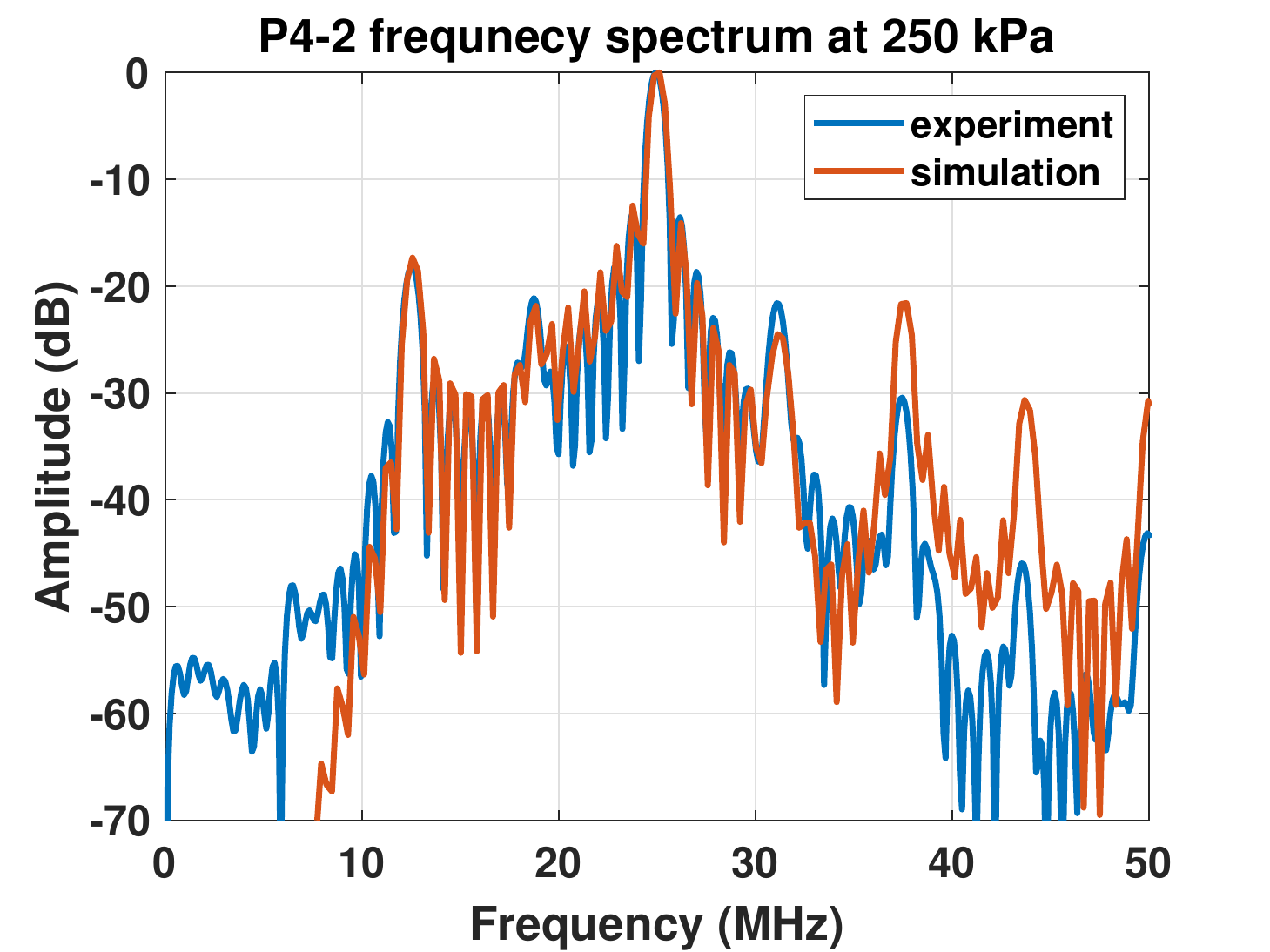}\\
		(j) \hspace{4cm}(k)\hspace{4cm}(l)\\
		\includegraphics[scale=0.3]{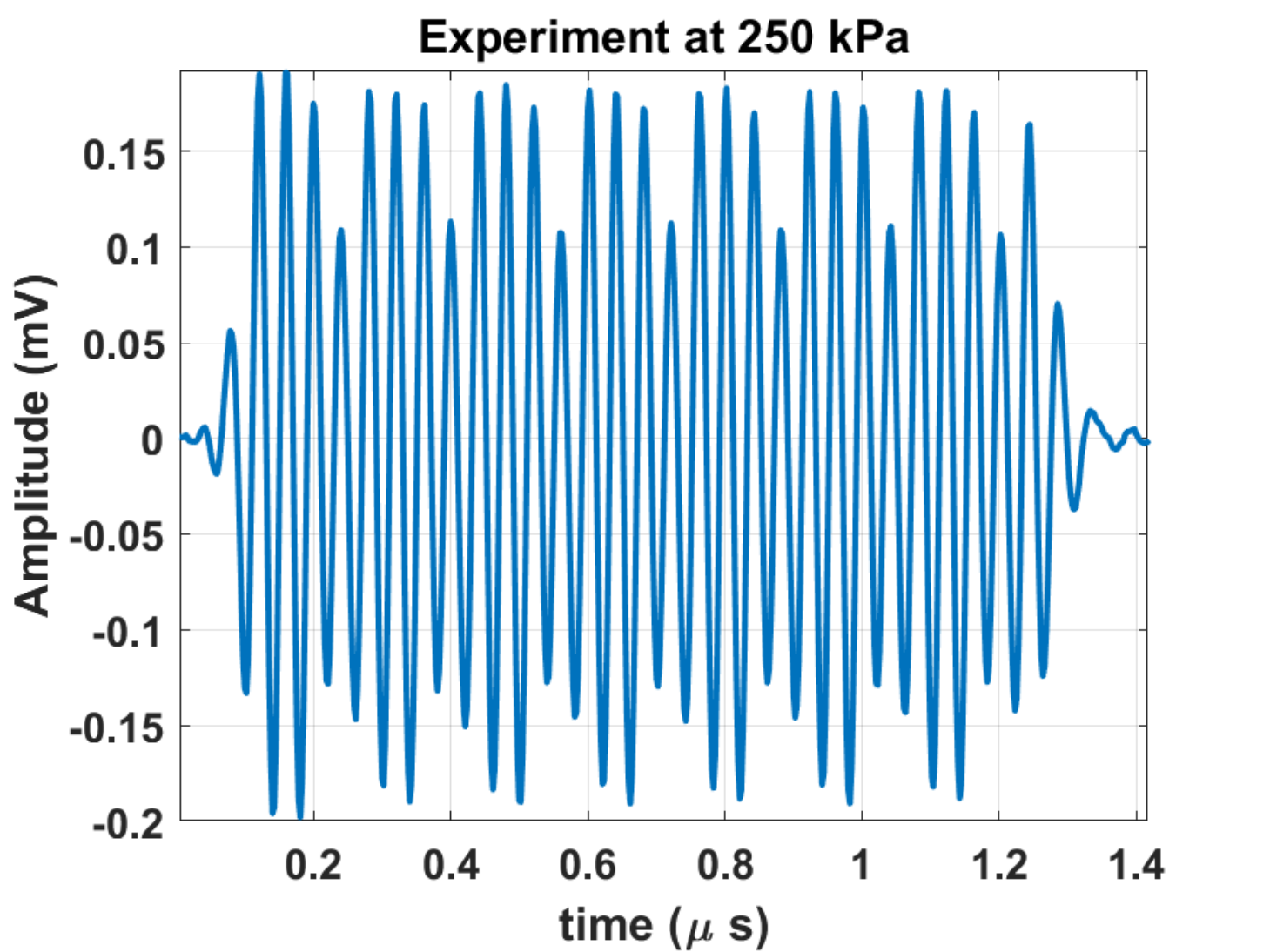}\includegraphics[scale=0.3]{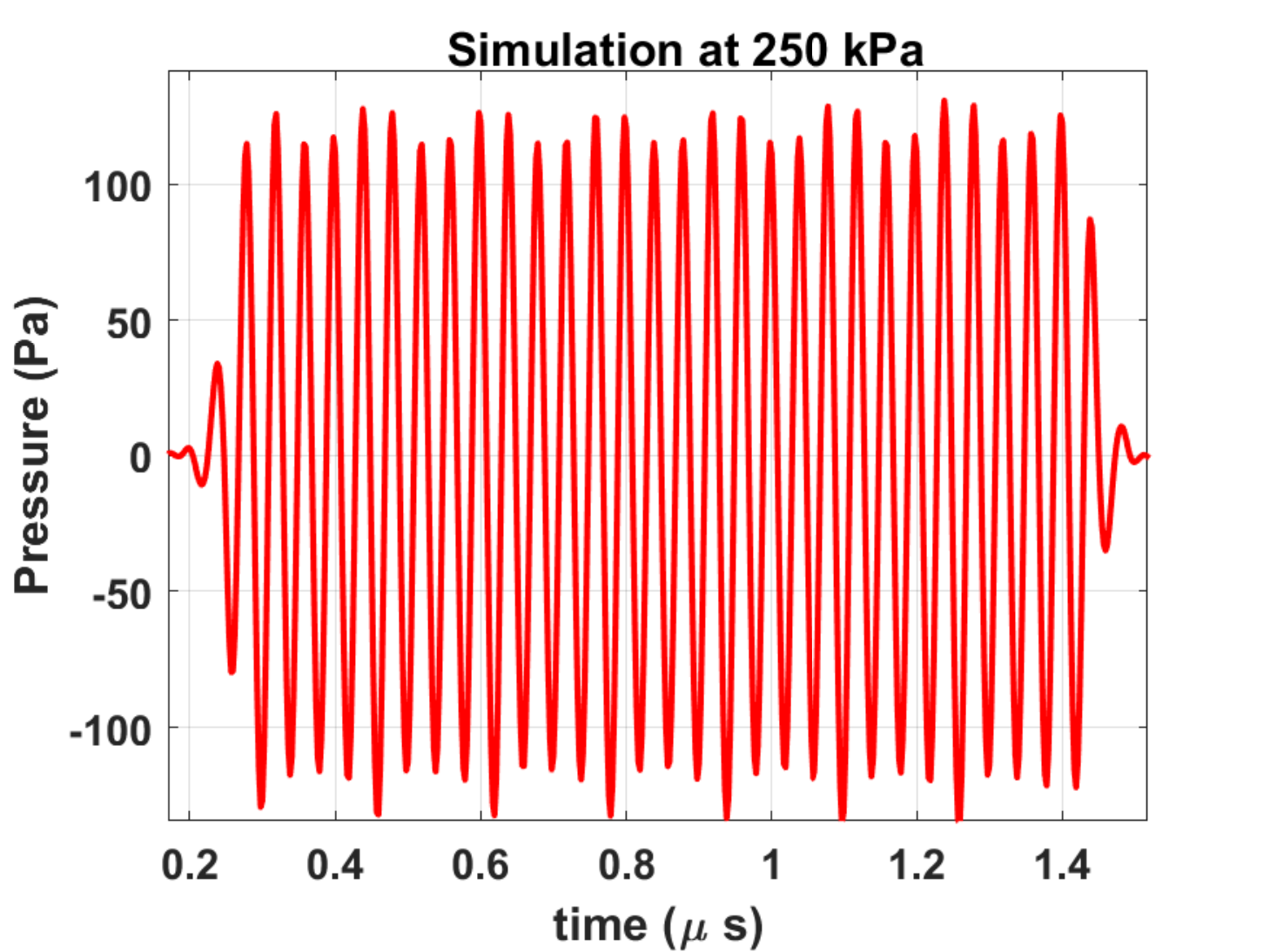}	\includegraphics[scale=0.3]{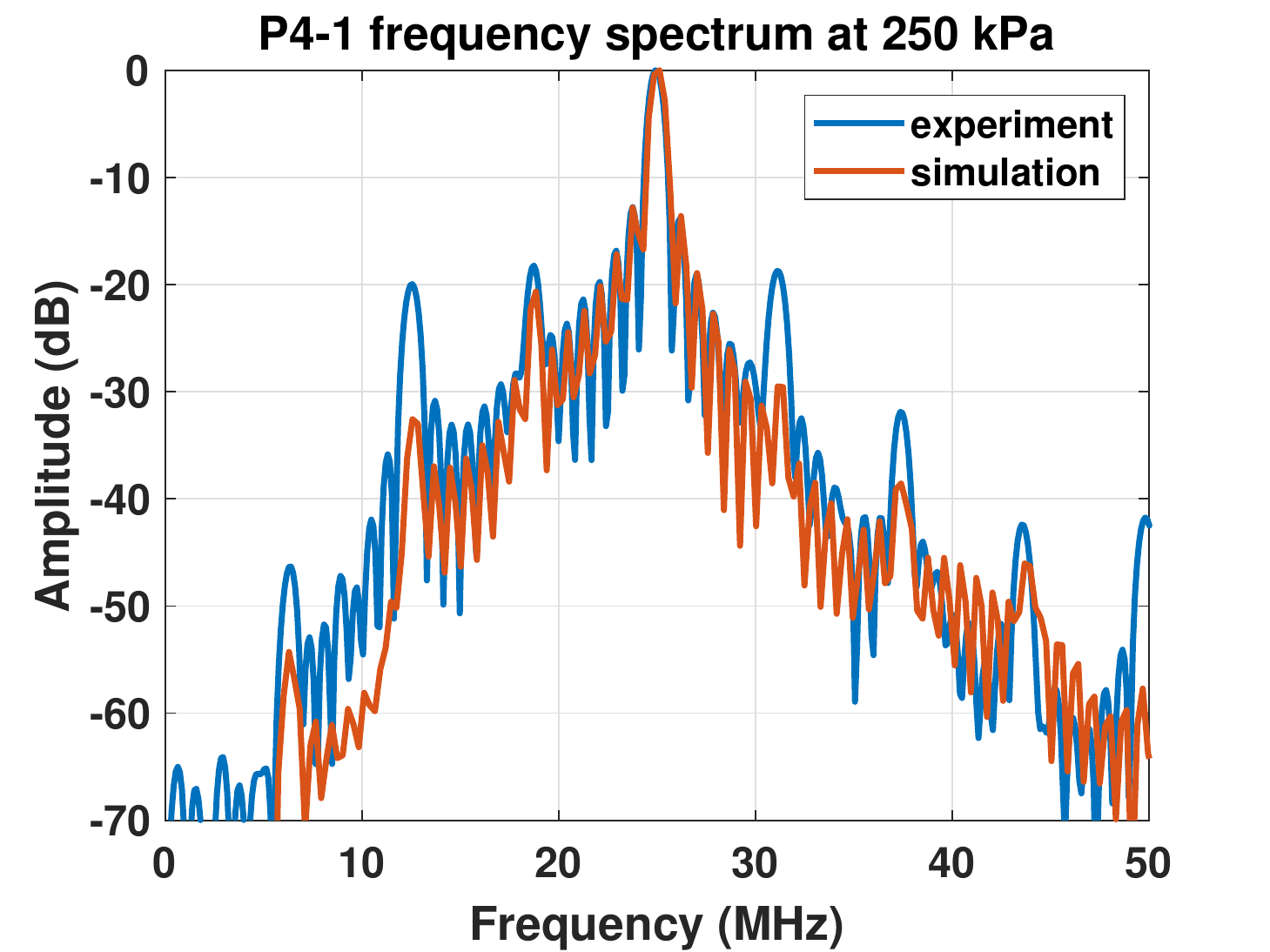}\\
		(m) \hspace{4cm}(n)\hspace{4cm}(o)\\
		\caption{ Demonstration of 5 main oscillation regimes acquired experimentally (blue) and simulated (red) choosing MB sizes based on the feature similarity in Fig. 3f. Representative experimental data and simulations of: 1st row P1, 2nd row P2, 3rd row P3, 4th row P4-2 and 5th row P4-1.}
	\end{center}
\end{figure*}
In experiments at $P_a\approx250 kPa$ and $f=25 MHz$ we observed 5 main types of backscattered signals in the data collected from single MB  events. A  representative of each category is shown in Fig. 5a (P1), Fig. 5d (P2), Fig. 5g (P3) , Fig. 5j (P4-2) and Fig. 5m (P4-1). The results of the numerical simulations are presented in the second column and the frequency spectrum of the experimental signals and the numerical simulations are plotted in the third column (blue:experiments, red:simulations). Numerical simulations are for the $Definity^{\textregistered}$ MBs with $\sigma(R_0)=0.072 N/m$ with the corresponding sizes chosen from the bifurcation diagram (Fig. 3f) to match the observed behavior in the experiments.\\
Fig. 5a displays a typical P1 signal observed in experiments. The calculated $P_{sc}$ for a 2 $\mu m$ $Definity^{\textregistered}$ MB  is displayed in Fig. 5b (in red color for distinction) and the power spectrum of the signals in Fig. 5a and 5b are shown in Fig. 5c. The scattered pressure has one maximum and the frequency spectrum has a peak at 25 MHz.\\ A representative signal of the P2 oscillations is displayed in the second row of Fig. 5. Both experimental and simulated (initial size of $0.955 \mu m$) signals have two maxima revealing a P2 oscillation regime. The power spectra in Fig. 5f  consist of a SH peak at 12.5MHz and a 3/2 UH peak at 32.5 MHz.\\ A representative of the P3 signal is shown in the third row of Fig. 5. The experimental and simulated (initial size of $1.39 \mu m$) signals have 3 maxima and the order of the maxima are consistent between experiments and simulations. The power spectra in Fig. 5i show a good agreement between experiments and simulations with SHs at (1/3 order) 8.33 MHz, (2/3 order) 16.66 MHz and  UHs at (4/3 order) 33.33 MHz and (5/3 order) 41.66 MHz.\\ 
\begin{figure*}
	\begin{center}
		\includegraphics[scale=0.4]{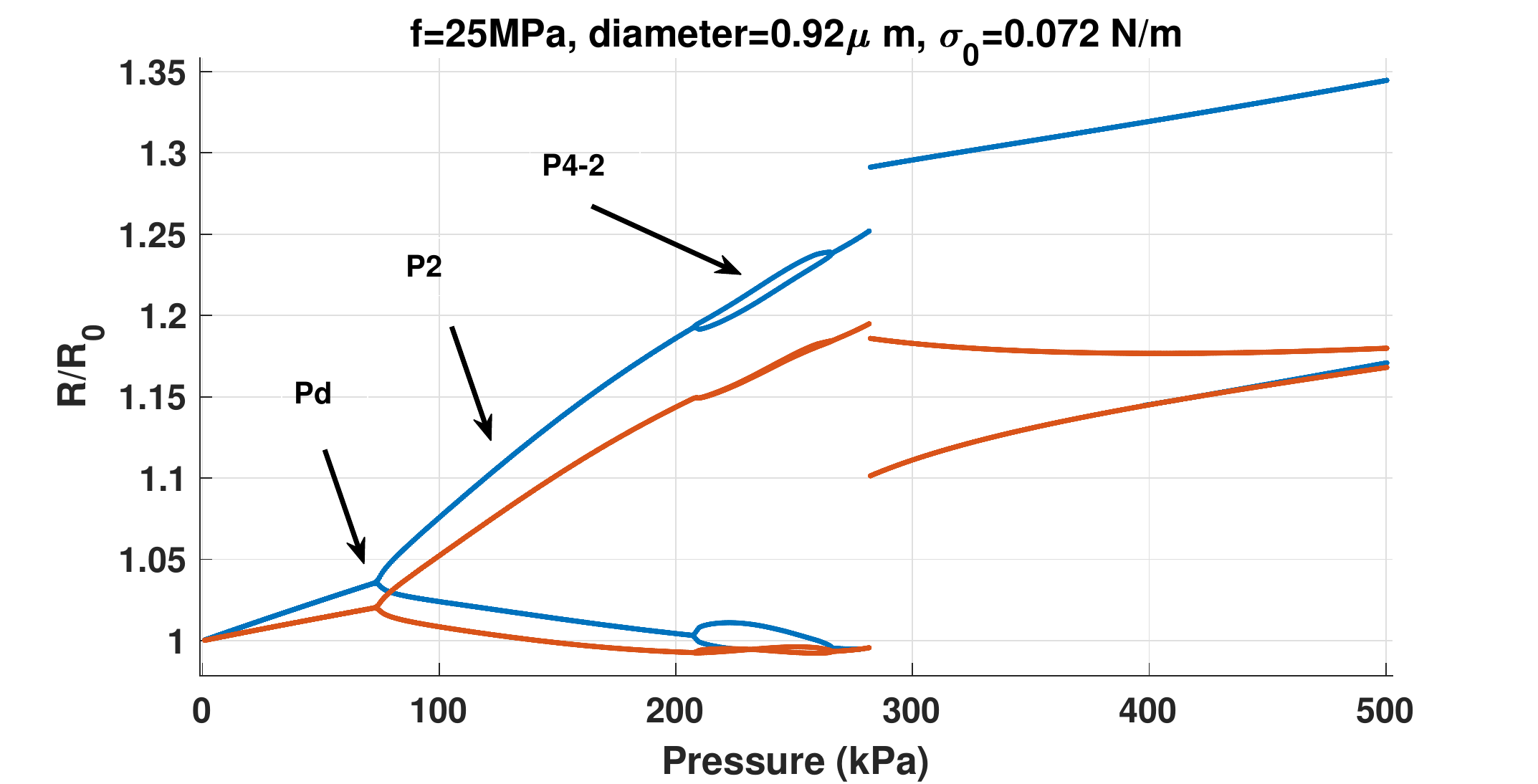}\\
		(a)\\
		\includegraphics[scale=0.4]{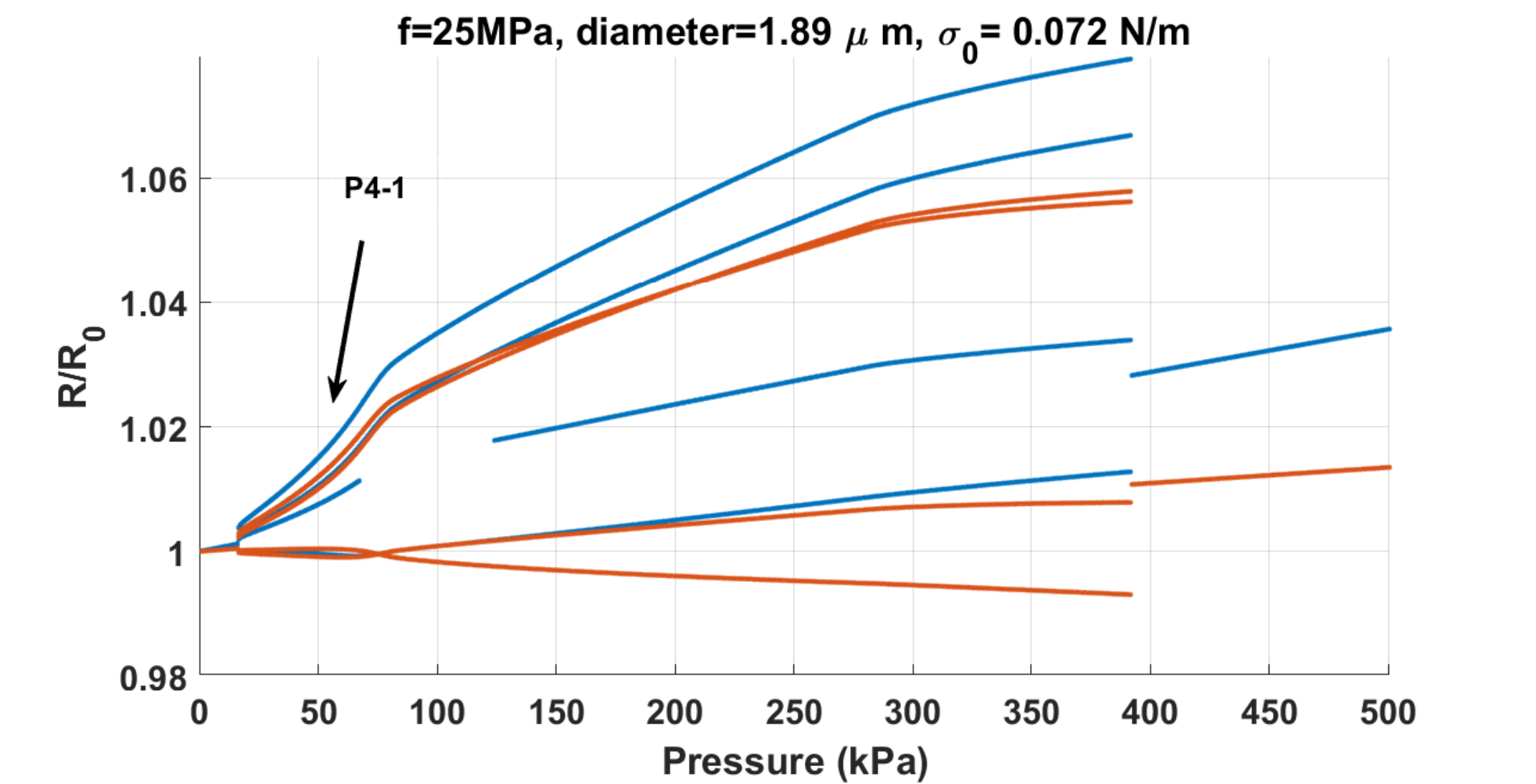}\\
		(b)\\
		\caption{The bifurcation structure of the $\frac{R}{R_0}$ of the MB as a function of pressure for excitation with $f=25 MHz$  for a lipid MB with $\sigma(R_0)=0.072 N/m$ and a diameter of : a) 0.92 $\mu m$ $\&$ b) 1.89 $\mu m$}
	\end{center}
\end{figure*}
P4-2 oscillations are shown in the 4th row of Fig. 5. There is a good agreement between the experimental and the simulated signals (initial size of $0.92 \mu m$). Both signals have 4 peaks in two envelopes and each envelope repeats itself once every two acoustic cycles. In each envelope there are two peaks and the peaks repeat themselves in an amplitude order of (largest, small, large, smallest). The frequency spectra of the signals are shown in Fig. 5l. There are 3 SHs at (1/4 order) 6.25 MHz, (1/2 order) 12.5 MHz and (3/4) order at 18.75 MHz. The 1/2 order SH is the strongest detected SH and due to the weakness of the 1/4 SH this peak is hardly detectable. This is because the transducer sensitivity drops sharply away from the center frequency and especially below 12.5 MHz (transducer bandwidth is 100$\%$). While the numerically simulated $P_{sc}$ in the absence of convolution with transducer response had a clear peak at 6.25 MHz, however, after the signal is convolved with the transducer response, the signal drops below the noise level of -70 dB in our experiments.\\ The last row of Fig. 5 depicts the case of the P4-1 oscillations. Simulations are for a MB with initial size of $1.89 \mu m$. The signals have one envelope with  4 maxima that repeats itself once every 4 acoustic cycles. Amplitudes repeat themselves in the order of smallest, largest, large and small. Both experimental and simulated signals demonstrate the same pattern of peaks and their orders. The power spectra in Fig. 5o shows a good agreement between the orders of the SHs and their locations.   There are 3 SHs at (1/4 order) 6.25 MHz, (1/2 order) 12.5 MHz and (3/4 order) at 18.75 MHz. The 3/4 order SH is the strongest detected SH. It should be noted that 1/4 order SH is the strongest peak in the calculated $P_{sc}$ in the absence of convolution with transducer response (see Fig. 7f). Due to the reduced sensitivity of the transducer at 6.25 MHz, the detected strength of the 1/4 order SH diminishes strongly and it drops below all the other SHs. \\  
\begin{figure*}
	\begin{center}
		\includegraphics[scale=0.3]{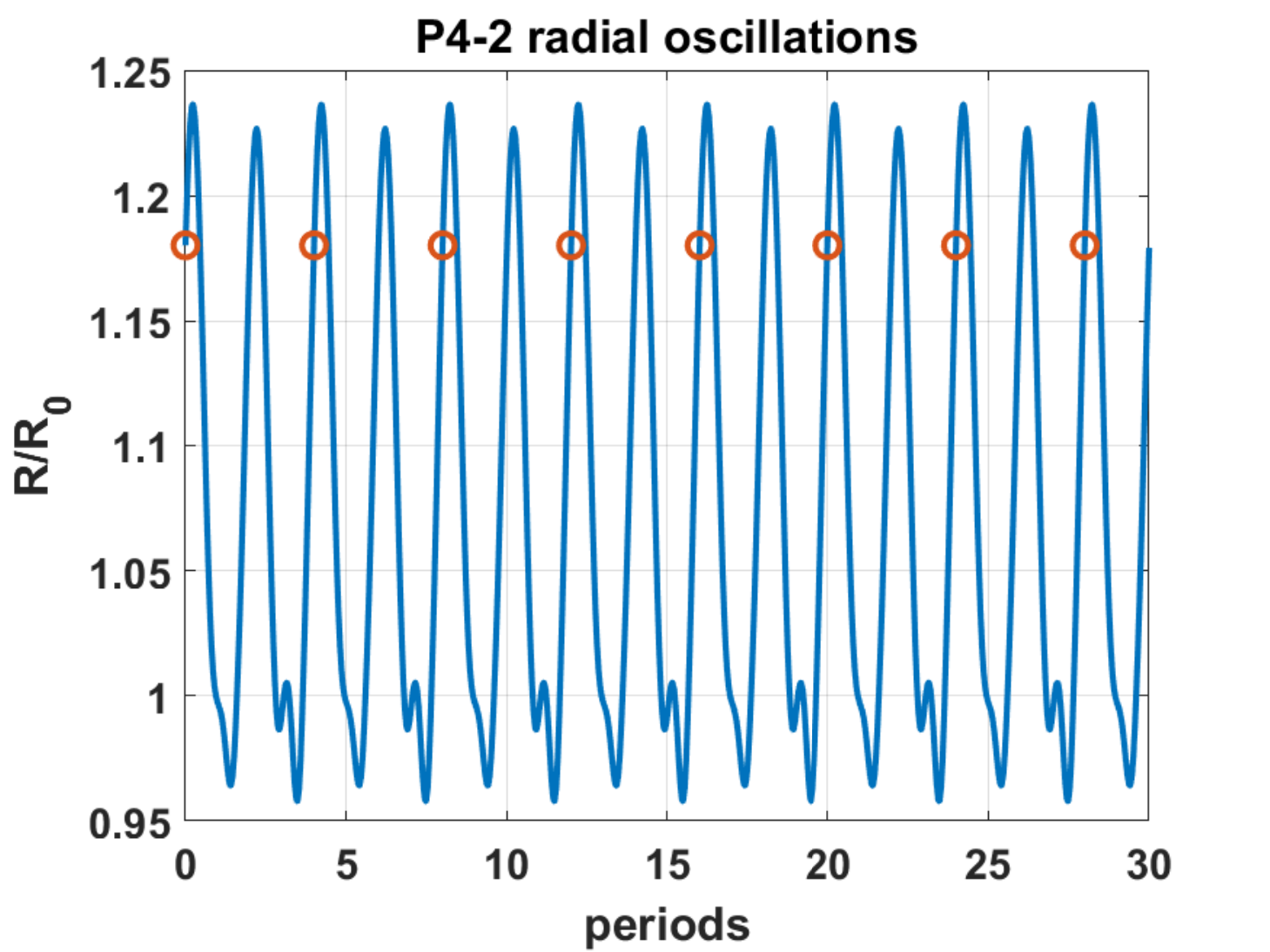}\includegraphics[scale=0.3]{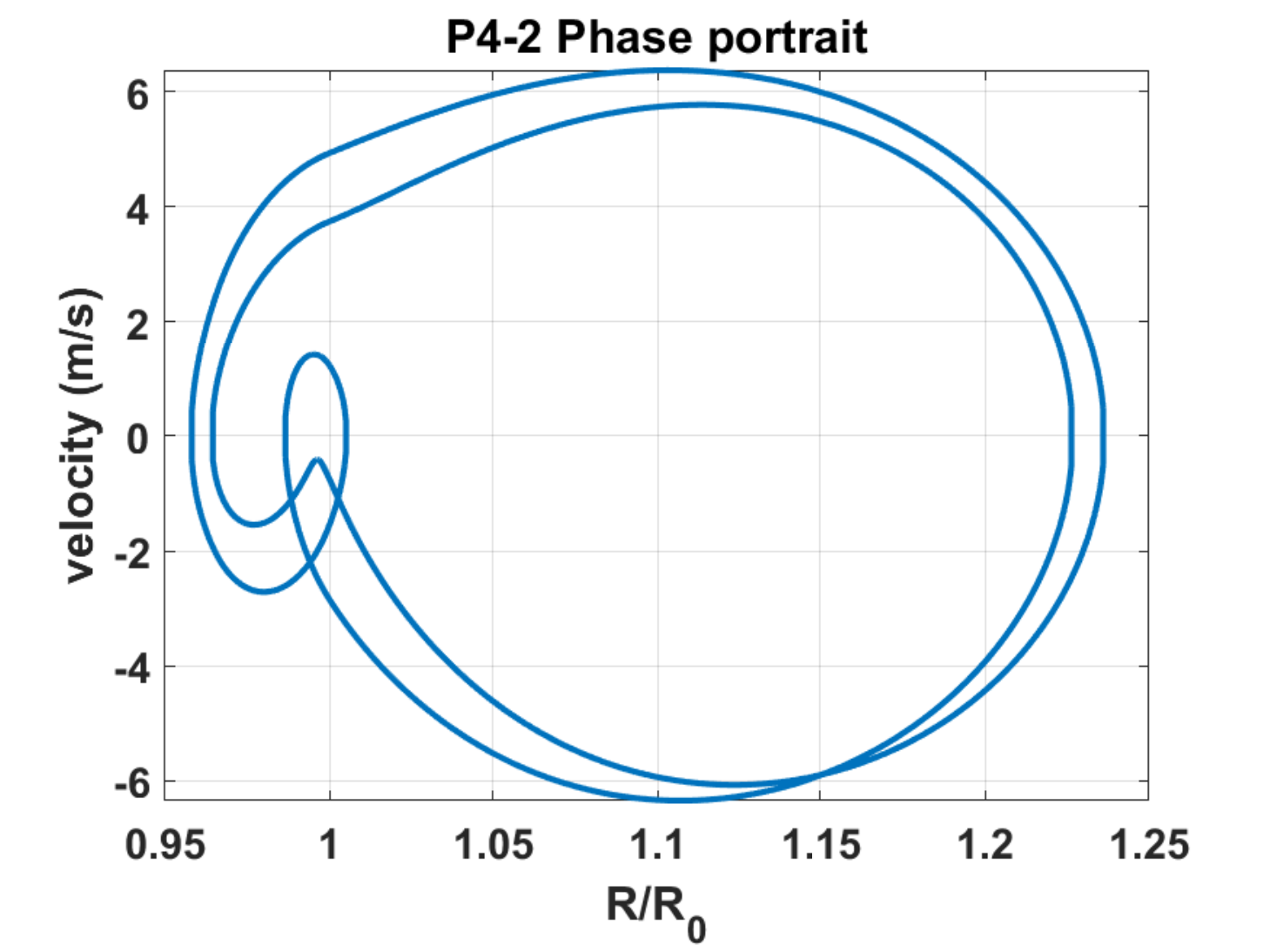}	\includegraphics[scale=0.3]{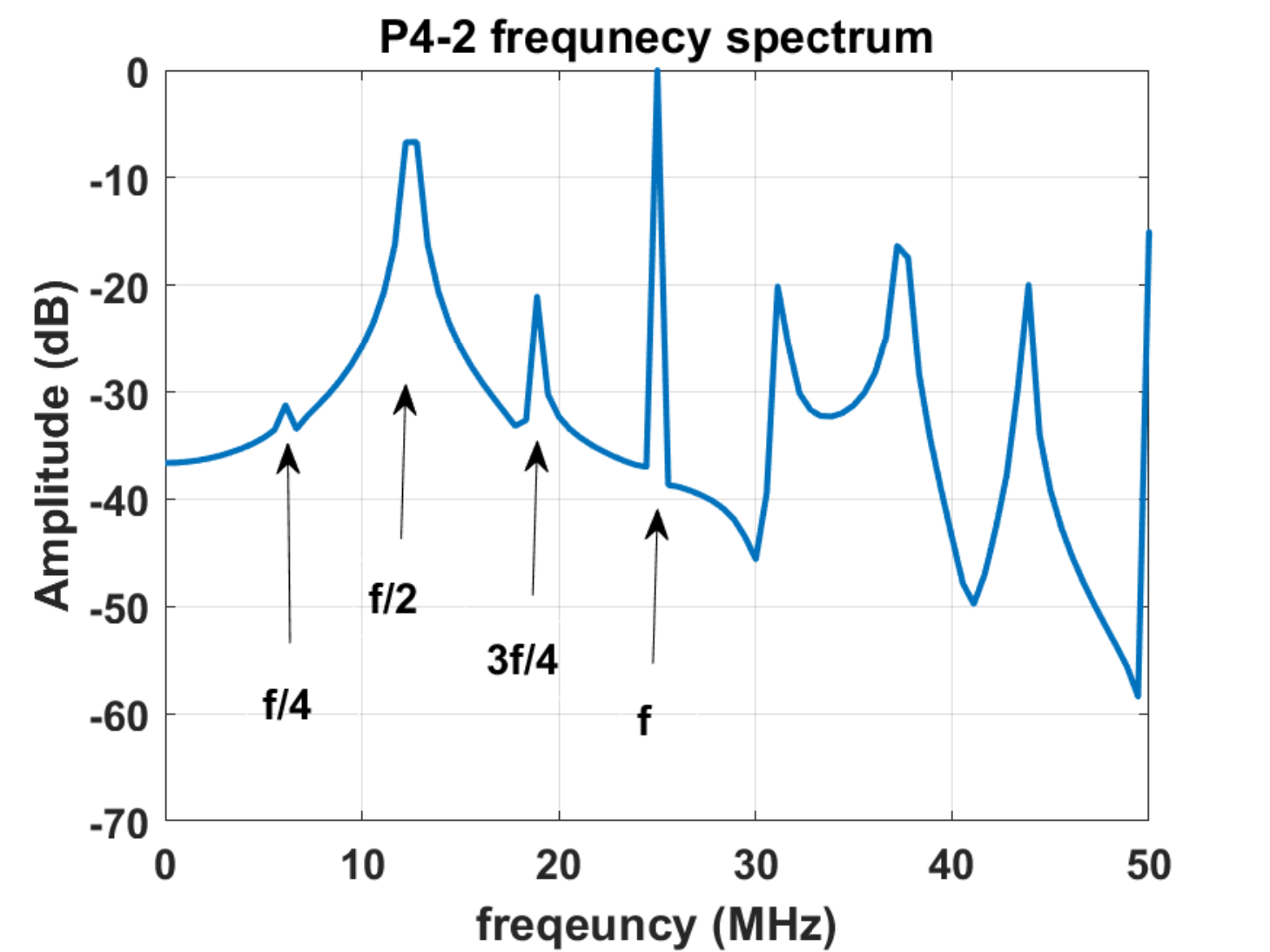}\\
		(a) \hspace{4cm}(b)\hspace{4cm}(c)\\
		\includegraphics[scale=0.3]{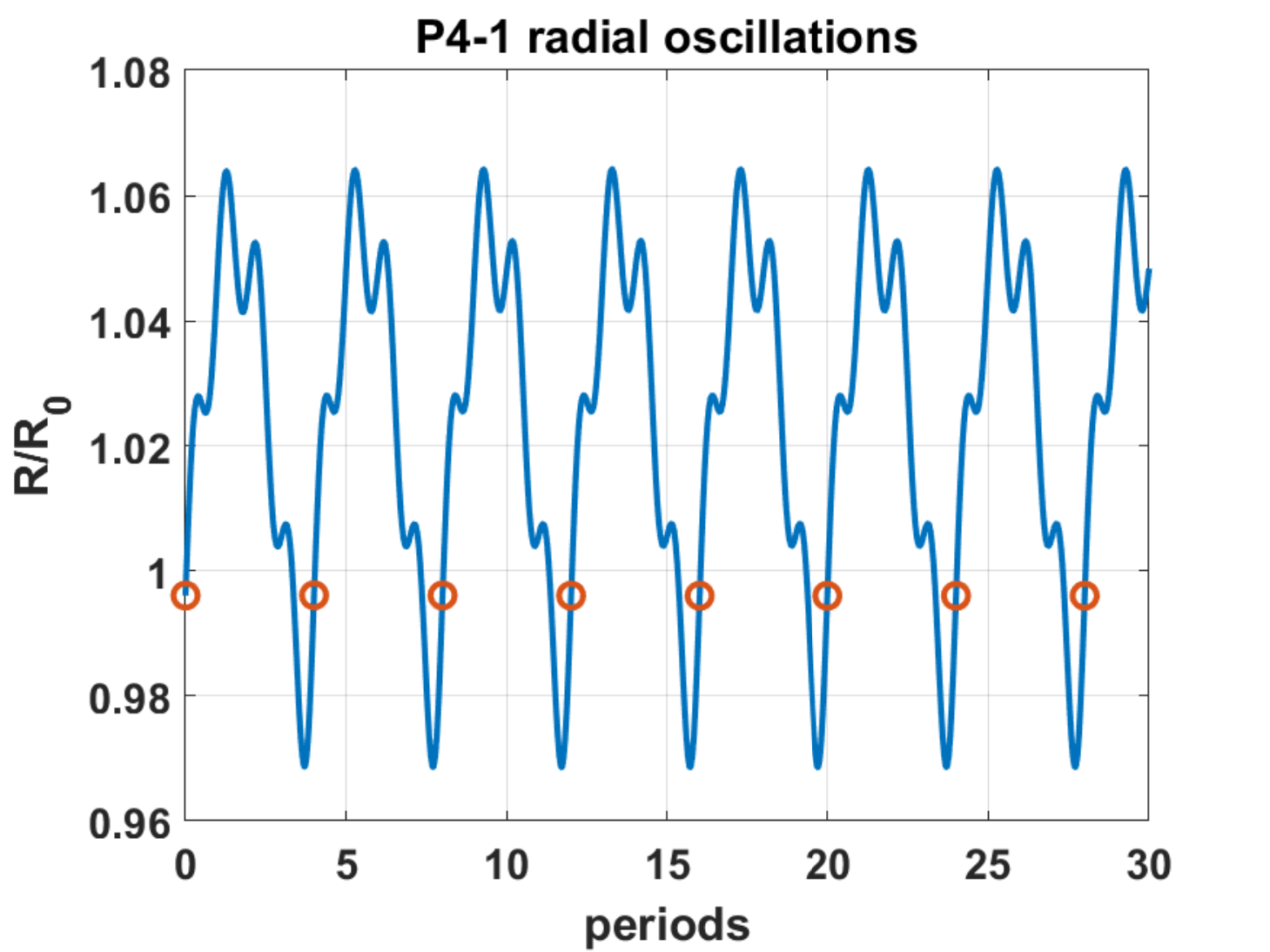}\includegraphics[scale=0.3]{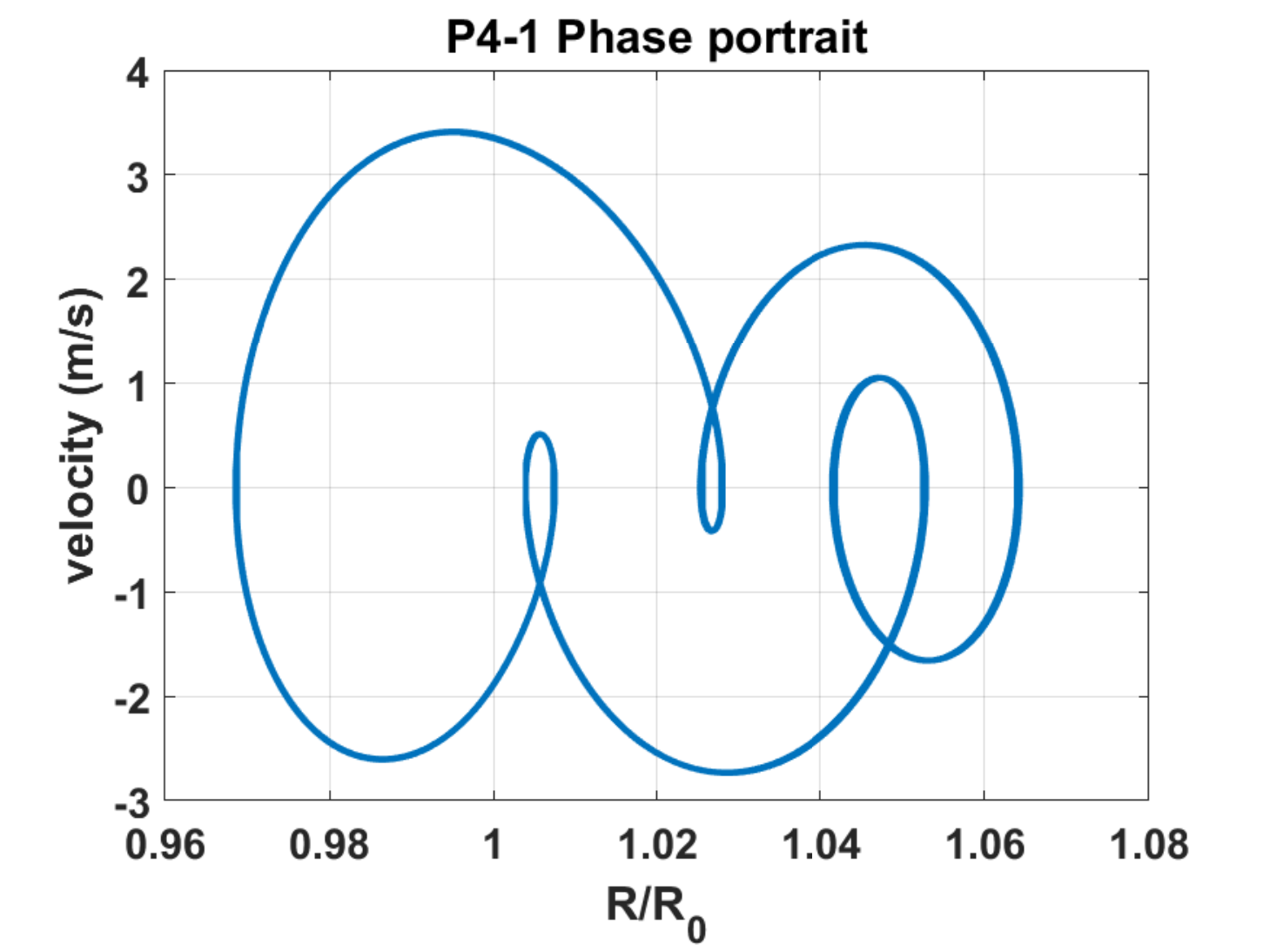}	\includegraphics[scale=0.3]{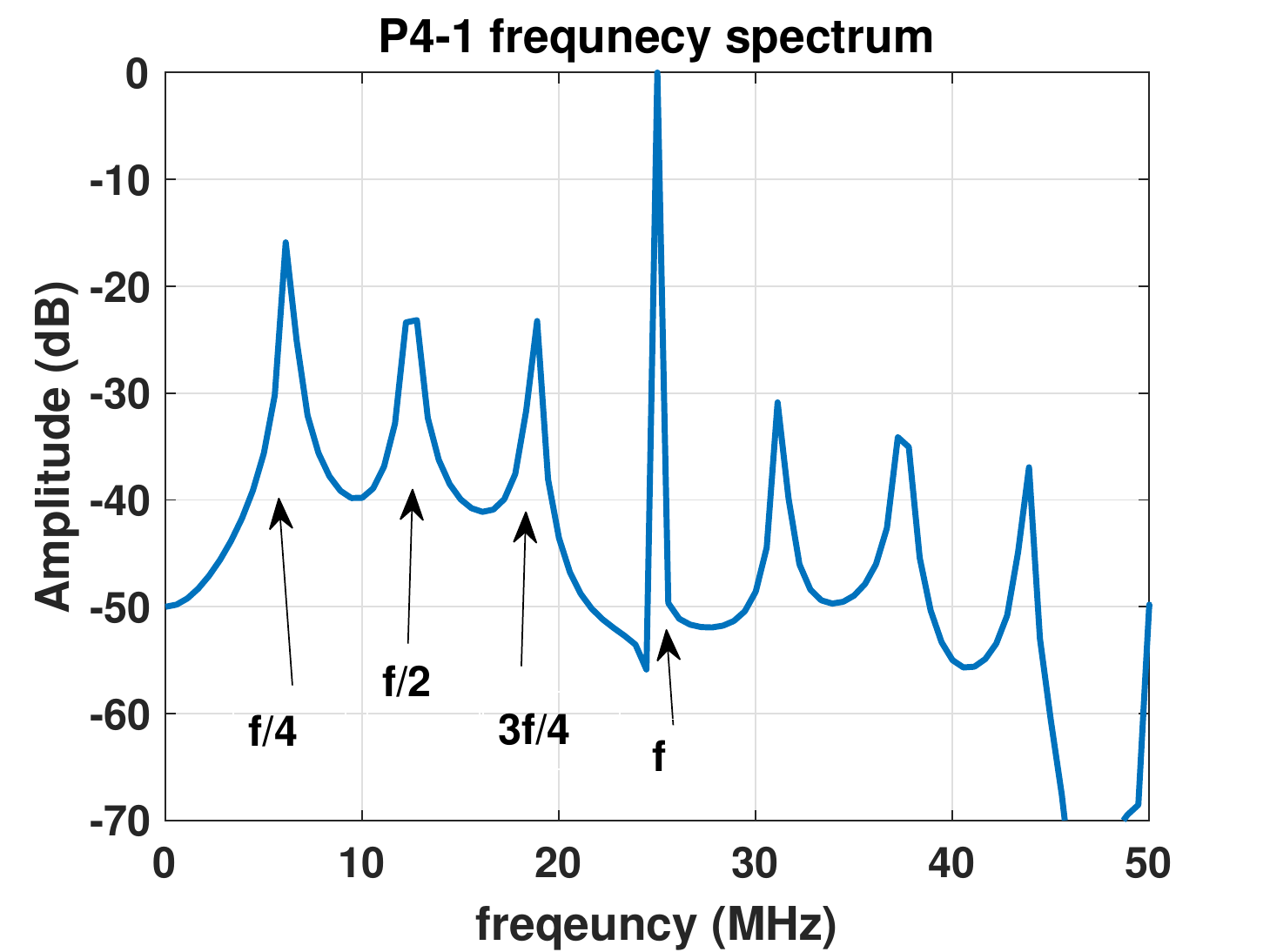}\\
		(d) \hspace{4cm}(e)\hspace{4cm}(f)\\
		
		\caption{ Characteristics of the two P4 oscillations identified: a) P4-2 radial oscillations, b) P4-2 phase portrait, c) power spectrum of the P4-2 $P_{sc}$, d)P4-1 radial oscillations, e)P4-2 phase portrait and f) power spectrum of P4-2 $P_{sc}$. Here $P_{sc}$ is not convolved with the transducer response. (red circles shows the location of the R every 4 acoustic cycles)}
	\end{center}
\end{figure*}
\subsubsection{Difference between P4-1 and P4-2}
Fig. 6a shows the bifurcation structure of a 0.92 $\mu m$ $Definity^{\textregistered}$ MB as a function of pressure when $f=25 MHz$. At low pressures there are linear oscillations with period doubling (Pd) at $\approx 70 kPa$. P2 oscillations undergo further Pd to P4-2 oscillations at $\approx 210 kPa$. The process of P4-2 generation and disappearance is through a bubbling bifurcation. In case of the  $1.89 \mu m$ $Definity^@$ (Fig. 6b), P4 oscillations are generated through a direct period quadrupling via a saddle node bifurcation similar to \cite{26}. This is the reason why we named this a P4-1 oscillations. Models for uncoated MBs or coated MBs with pure viscoelastic behavior predict very high pressures for the generation of P4-1 oscillations; however, here we show, for the first time, that the dynamic variation of the shell elasticity including buckling and rupture enhances the generation of the P4-1 oscillations at very low acoustic pressures ( $P_a\approx16.6 kPa$ in Fig. 6b).\\
Fig. 7 compares the radial oscillations, phase portraits and the power spectra of the $P_{sc}$ for both P4 oscillations at $f=25 MHz$ and $P_a=250 kPa$. P4-2 radial oscillations consist of two envelops, with each envelope having 2 maxima or one with 2 maxima and the other with a maxima and a critical point. These envelopes repeat themselves once every two acoustic cycles in Fig. 7a. The phase portrait of the P4-2 oscillations consists of a loop undergoing two internal loops with the largest loop undergoing another internal loop. The power spectrum depicts SHs with strength order of $1/2>3/4>1/4$. P4-1 oscillations in Fig. 7d have one envelope with 4 maxima which repeats itself once every 4 acoustic cycles. The phase portrait consist of a main loop that has undergone 3 bends to create 3 internal loops. The frequency spectrum of $P_{sc}$ depicts SHs in the strength order of $1/4>1/2>3/4$. It should be noted that due to the lower sensitivity of the transducer as we move away from central frequency, the strength order of the SHs that are detected in experiments were different. After, convolving the simulations results with the one way transducer response, experiments and simulations were in good agreement. To our best knowledge, this is the first time that the two types of P4 oscillations are detected experimentally and characterized numerically for a MB oscillator.    
\section{Discussion}
A MB oscillator is an extremely complex system that has beneficial applications in a wide range of fields including material science and sonochemistry \cite{59,60,61}, food technology \cite{62} underwater acoustics \cite{63,64} and medical ultrasound (ranging from imaging blood vessels \cite{65}, drug delivery \cite{51} to thrombolysis \cite{66} and the treatment of brain through intact skull \cite{49,67}). In addition to these important applications, the complex dynamical properties of the MB system make it a very interesting subject in the field of nonlinear dynamics. It is well known that an ultrasonically excited MB is a highly nonlinear oscillator. Due to the importance of the understanding of the MB behavior in several applications, numerous studies have employed the methods of nonlinear dynamics and chaos to study the complex behavior of the system. Pioneering works of \cite{20,61,68} have revealed several nonlinear and chaotic properties of the MB oscillations (both numerically and experimentally). Recent extensive studies on the nonlinear behavior of MBs in water \cite{23,24,25,26,27,43,44}, coated MBs \cite{26,27,42}, MBs in highly viscous media \cite{28,29,30,31,32,33}, MBs sonicated with asymmetrical driving acoustic forces \cite{33,34,35,36,37,38,39,40,41} and MBs in non-Newtonian fluids \cite{37} have revealed many nonlinear features in the MB behavior. Occurrence of P2, P3, P4-2, P4-1 and higher periods, as well as chaotic oscillations, has been demonstrated in these works. Moreover, the effect of nonlinear dynamics of MBs on the propagation of sound waves in a bubbly medium is under recent investigation\cite{26,69,70}.\\ Despite these studies that employed the methods of chaos physics to investigate the nonlinear dynamics of the uncoated and coated MBs with viscoelastic behavior, the effect of the lipid coating on the dynamics of the MB especially in the realm of nonlinear dynamics and chaos has not been systematically investigated.\\ 
In this study we investigated the bifurcation structure of the lipid coated MBs and used the numerical results to help interpret unique signals that we observed experimentally. In stark contradiction to the results of classical theory of uncoated MBs, and despite the increased damping of the coated MBs, lipid coated MBs exhibited higher order nonlinear behavior at low excitation amplitudes (shown here both experimentally and numerically). The numerical and experimental findings can be summarized as follows:\\ 
a- Theoretically, we have shown that even at pressures as low as $5 kPa$, $6th-2nd$ order SuHs, P4-2, P2, P3 P4-1 and chaotic regimes manifest themselves in the MB behavior. To our best knowledge the existence of higher order SHs and chaotic behavior at such low excitation amplitudes is first reported here.\\   
b- The initial surface tension of the MB plays a critical role in the enhanced nonlinear behavior. We have shown that the closer $\sigma(R_0)$ is to 0 (leading to buckling and compression only behavior) or to $\sigma_{water}$ (leading to shell rupture and expansion dominated behavior), the lower the excitation threshold for nonlinear behavior and the higher the order of non-linearity.\\
c- Despite the increased damping of the lipid coated MBs we show that, the MBs with surface tension $\geq0.062 N/m$ may have higher radial oscillation amplitude compared to the uncoated bubble.\\
d- We have experimentally shown that single  $Definity^{\textregistered}$ MBs, can exhibit, P2, P3, P4-2 and P4-1 oscillations at high frequencies (25 MHz) and low pressures (250 kPa). These results can not be predicted using conventional coated MB models (with pure viscoelastic behavior) and they even contradict predictions of uncoated MB models with less damping effects.\\
e-  Through numerical simulations of Marmottant model \cite{12} and visualization of the results using bifurcation diagrams we showed that $Definity^{\textregistered}$ MBs can exhibit enhanced nonlinear behavior. Using this model and assuming MBs with initial surface tension close to 0 N/m or $\sigma_{water}$ could be used to explain experimental observations of higher order nonlinear oscillations\\
f- The 5 main regimes of oscillations were identified as P1, P2, P3, P4-2 and P4-1. Simulation results of the scattered pressure were in good agreement with experimental observations both in terms of the shape of the amplitude versus time signal and also its frequency content.\\
g- For the first time, the two different P4 oscillations of the MB system were identified and characterized experimentally and numerically.  P4-2 oscillations are the result of two consecutive well known period doublings while P4-1 oscillations occur through a single period quadrupling via a saddle node bifurcation. The distinct features of the signal shapes and their unique frequency spectrum were identified both experimentally and numerically. P4-1 oscillations require larger MBs compared to P4-2 oscillations.\\
Previous studies have shown that lipid coated MBs can exhibit 1/2 order subharmonic oscillations even when the excitation amplitude is low ($<$30 kPa \cite{1,2,3}) where such low pressure thresholds are below the thresholds expected even for uncoated free MBs \cite{4,5}. The low pressure threshold for SH emissions has been attributed to the buckling of the coating and compression only behavior \cite{1}. Compression dominated oscillations \cite{6} occur when the coating buckles and the effective surface tension on the MB drops to values close to zero. In such an instance, the MB compresses more than it expands. In addition to compression only behavior, lipid coating may also result in expansion dominated behavior where the MB expands more than it compresses \cite{11,13}. Expansion-dominated behavior occurs when the shell ruptures. This effect was used to explain the enhanced non-linearity at at higher frequencies (25 MHz) \cite{11,72}. Theoretical analysis of the Marmottant model for lipid coated MBs \cite{12} by Prosperetti \cite{4} attributed the lower SH threshold of the lipid MBs to the variation in the mechanical properties of the coating in the neighborhood of a certain MB radius (e.g. occurrence of buckling). In this work we show that there is a symmetry for enhanced non-linearity in the bifurcation structure of the $\frac{R}{R_0}$ of the MB as a function of $\sigma(R_0)$. Both buckling and rupture can be responsible for enhanced non-linearity, where the closer the $\sigma(R_0)$ to the buckling state (0 N/m) or rupture threshold (0.072 N/m), the lower the excitation threshold required for the generation of nonlinear oscillations. Moreover, the closer the $\sigma(R_0)$ to these two limit values, the higher the order of the nonlinearity.\\
Using the estimated parameters for the $Definity^{\textregistered}$ MB  in \cite{54} and considering the shear thinning \cite{53}, the observed experimental behavior was only replicated for MBs with initial surface tension close to the two limit values of 0 and 0.072 N/m.  However, it should be noted that during the sonication of a polydisperse solution of lipid MBs different values in initial surface tension and coating properties (coating elasticity and viscosity) are expected. It is been reported that even for MBs of the same size, the lipid coating can be different from MB to MB and are shown to be heterogeneous for MBs smaller than 10 $\mu m$\cite{73,74}. Despite the better homogenity of lipid distribution in lipid coated MBs similar to $Definity^{\textregistered}$ \cite{73,74}, the small differences in the lipid distribution in the coating influences the effective coating properties, thus changing the MB response \cite{75,76,77}. Moreover, its shown that the coating elasticity and coating viscosity changes with the MB size \cite{78,79,80}. Despite assuming the same coating properties for all MB  sizes in this work, we were still be able to replicate the peculiar higher order nonlinearities in experiments. Moreover, we used the simplest model for lipid coated MBs and we neglected the possible stiffness softening \cite{81} or higher viscoelastic effects. Implementation of these effects are outside of the focus of this study but can be used to better characterizing the coating. In addition, simulation results only implemented a monofreqeuncy ultrasound source, and the effects of nonlinear propagation of sound waves in the medium are neglected. The generation of the SHs and UHs were not due to the nonlinear propagation of waves as nonlinear propagation manifests itself through generation of only harmonics.\\
Effects of the shape oscillations on the MBs response were also neglected in this paper. Holt and  Crum observed significant effects of shape oscillations on the nonlinear behavior of the larger MBs with initial radii between $20 \mu m<R_0< 100 \mu m$ \cite{82}. Versluis et al. \cite{83} using high speed optical observations identified time-resolved shape oscillations of mode n= 2 to 6 in the behavior of single air bubbles with radii between $10 \mu m$ and 45 $\mu m$. \cite{83} concludes that close to resonance, bubbles were found to be most vulnerable toward shape instabilities. The effect of non-spherical bubble oscillations on nonlinear bubble behavior is studied in \cite{84} through GPU accelerated large parameter investigations. The active cavitation threshold has been shown to depend on the shape instability of the bubble \cite{84}. \cite{84} also shows that shape instability can affect the subharmonic threshold and nonlinear behavior of bubbles. Nonspherical oscillations of ultrasound contrast agent coated MBs are investigated in \cite{85} through high speed optical observations. They
showed that non-spherical bubble oscillations are significantly present in medically relevant ranges of bubble radii and applied acoustic pressures. Non-spherical oscillations develop preferentially at the resonance radius and may be present during SH oscillations \cite{85}. Thus, for a more accurate modeling of the MB oscillations, more sophisticated theoretical modeling of bubble coating, accounting for membrane shear and bending is required \cite{85}. Recently Guédra et al. \cite{86} showed that at sufficiently large pressure amplitudes, energy transfer from surface to volume oscillations may trigger subharmonic spherical oscillations at smaller pressure amplitudes than predicted by models of spherical only bubble oscillations. Guédra and Inserra  studied the nonlinear interactions between spherical and nonspherical modes in \cite{87}. They showed that bubble
shape oscillations could be sustained for excitation amplitudes below the classical parametric threshold.
Experimentally, nonlinear interactions between the spherical, translational, and shape oscillations of micrometer-size bubbles have been reported in detail in \cite{88}. Liu et al \cite{POF2} numerically studied the shape oscillations of encapsulated bubbles. They showed that, in case of very small encapsulated microbubbles, the shape oscillation is less likely to occur since the surface tension suppresses the developments of nonspherical shape modes. Their model however, does not take into account the dynamic variations in the effective surface tension that occurs in case of lipid coated bubbles.   Liu and Wang \cite{POF3} showed that shape modes  of  an  encapsulated microbubbles  in  a  viscous  Newtonian  liquid  are stable. In case of encapsulated bubbles with stiffness hardening or softening behavior, Tisiglifis and Plekasis \cite{POF4} numerically showed that parametric instability is possible and result in shape oscillations as a result of subharmonic or harmonic resonance. Generation of the shape modes for encapsulated bubbles exhibitng breakup and buckling have been nurmeically investigated for higher bubble oscillation amplitude \cite{POF5}. Implementation of the nonspeherical bubble oscillations is beyond the scope of current paper, however, it may help to achieve a better agreement between numerical simulations and experimental observations in case of larger bubbles (e.g. Figure 5m-o).\\
Since experimental study was done in very dilute solutions of MBs to record single scattering events, bubble-bubble interaction was not included in the model. Bubble-bubble interaction have been shown to lower the threshold of subharmonic emissions and chaos \cite{44,89,90,91}. However even at very high concentrations \cite{44} the changes to the pressure threshold of subharmonics is minimal compared to the signification decrease in the onset of higher order nonlinearities when the shell undergoes buckling and rupture.\\
 Generation of higher order SHs at low pressures may have potential in high resolution SH imaging due to their higher frequencies, higher contrast to tissue and signal to noise ratio. A SH of order 2/3 or 3/4 can be detected more effectively by the transducer as they are closer to the transducer center frequency when compared to 1/2 order SHs. Moreover, the higher scattered pressures, faster oscillations and the lower frequency contents of the oscillations of the higher order SHs may enhance the nondestructive shear stress on cells for enhanced drug delivery or in cleaning applications. Mixing applications are another category of applications that can take advantage of higher order SHs at high frequencies.  
\section{Conclusion}
We have shown experimentally and for the first time that higher order SHs (e.g. 1/3,1/4,..) can be generated in the oscillations of lipid coated MBs when insonated at high frequencies and low excitation amplitudes. The bifurcation structure of a simple model of lipid coated MBs were studied as function of frequency and size to explain the experimental observations. We showed that compression only behavior or expansion dominated oscillations due to buckling and rupture of the coating and dynamic variation of the effective surface tension can explain the observed enhanced non-linearity in MBs oscillations.\\
\textbf{Acknowledgments}\\
The work is supported by the Natural Sciences and Engineering Research Council of Canada (Discovery Grant RGPIN-2017-06496), NSERC and the Canadian Institutes of Health Research ( Collaborative Health Research Projects ) and the Terry Fox New Frontiers Program Project Grant in Ultrasound and MRI for Cancer Therapy (project $\#$1034). A. J. Sojahrood is supported by a CIHR Vanier Scholarship and Qian Li is supported by NSF CBET grant $\#$1134420.

\end{document}